\documentclass[twocolumn,preprintnumbers,amsmath,aps]{revtex4}
\usepackage{graphicx}
\usepackage{dcolumn}
\usepackage{bm}
\usepackage{color}
\usepackage{multirow}
\usepackage{epstopdf}
\begin{document}
\newcommand{\kvec}{\mbox{{\scriptsize {\bf k}}}}
\newcommand{\qvec}{\mbox{{\scriptsize {\bf q}}}}
\def\eq#1{(\ref{#1})}
\def\fig#1{Fig.\hspace{1mm}\ref{#1}}
\def\tab#1{Tab.\hspace{1mm}\ref{#1}}
\title{Anomalously high value of Coulomb pseudopotential for the $\rm H_{5}S_{2}$ superconductor}
\author{Ma{\l}gorzata Kostrzewa $^{\left(1\right)}$}
\email{malgorzata.kostrzewa@ajd.czest.pl}
\author{Rados{\l}aw Szcz{\c{e}}{\'s}niak $^{\left(1, 2\right)}$}
\author{Joanna K. Kalaga$^{\left(3\right)}$}
\author{Izabela A. Wrona $^{\left(1\right)}$}
\affiliation{$^{\left(1\right)}$ Institute of Physics, Jan D{\l}ugosz University in Cz{\c{e}}stochowa, Ave. Armii Krajowej 13/15, 42-200 
Cz{\c{e}}stochowa, Poland}
\affiliation{$^{\left(2\right)}$ Institute of Physics, Cz{\c{e}}stochowa University of Technology, Ave. Armii Krajowej 19, 42-200 Cz{\c{e}}stochowa, Poland}
\affiliation{$^{\left(3\right)}$ Quantum Optics and Engineering Division, Faculty of Physics and Astronomy, University of Zielona G{\'o}ra, Prof. Z. Szafrana 4a, 65-516 Zielona G{\'o}ra, Poland}
\date{\today} 
\begin{abstract}
The ${\rm H_{5}S_{2}}$ and ${\rm H_{2}S}$ compounds are the two candidates for the low-temperature phase of compressed sulfur-hydrogen system. We have shown that the value of Coulomb pseudopotential ($\mu^{\star}$) for ${\rm H_{5}S_{2}}$  
($\left[T_{C}\right]_{\rm exp}=36$~K and  $p=112$~GPa) is anomalously high. The numerical results give the limitation from below to 
$\mu^{\star}$ that is equal to $0.402$ ($\mu^{\star}=0.589$, if we consider the first order vertex corrections to the electron-phonon interaction). 
Presented data mean that the properties of superconducting phase in the ${\rm H_{5}S_{2}}$ compound can be understood within the classical framework 
of Eliashberg formalism only at the phenomenological level ($\mu^{\star}$ is the parameter of matching the theory to the experimental data). 
On the other hand, in the case of ${\rm H_{2}S}$ it is not necessary to take high value of Coulomb 
pseudopotential to reproduce the experimental critical temperature relatively well ($\mu^{\star}=0.15$). In our opinion, ${\rm H_{2}S}$ is mainly responsible for the observed superconductivity state in the sulfur-hydrogen system at low temperature.

\vspace*{0.2cm}

\vspace*{0.2cm}
\end{abstract}
\maketitle
%

\section{INTRODUCTION}
The Eliashberg formalism \cite{Eliashberg1960A} allows for the analysis of classical phonon-mediated superconducting state at the quantitative level. 
The input parameters to the Eliashberg equation are: the spectral function, otherwise called the Eliashberg function ($\alpha^{2}F\left(\omega\right)$) that is modeling the electron-phonon interaction \cite{Carbotte1990A, Carbotte2003A}, and the Coulomb pseudopotential ($\mu^{\star}$), which is responsible for the depairing electron correlations \cite{Morel1962A, Bauer2012A}. Usually, the Eliashberg function is calculated using the DFT method 
({\it e.g.} in the Quantum Espresso package \cite{Baroni1986A, Giannozzi2009A}). The value of Coulomb pseudopotential is selected in such way that the critical temperature, determined in the framework of Eliashberg formalism, would correspond to $T_{C}$ from the experiment. Sometimes $\mu^{\star}$ is tried to be calculated from the first principles, however, this is the very complex issue and it is rarely leading to correct results \cite{Kim1993A}. 

It should be noted that the value of Coulomb pseudopotential should not exceed $0.2$. For higher values, $\mu^{\star}$ cannot be associated only with the depairing electron correlations. In the case at hand, the quantity $\mu^{\star}$ should be treated only as the parameter of fitting the model to the experimental data, which means that the Eliashberg theory becomes practically the phenomenological approach.

The phonon-mediated superconducting state in the compressed ${\rm H_{2}S}$ with high value of critical temperature was discovered 
by Drozdov, Eremets, and Troyan in 2014 \cite{Drozdov2014A} (see also the paper \cite{Drozdov2015A}). The experimental observation of superconductivity in the compressed dihydrogen sulfide was inspired by the theoretical prediction made by Li {\it et al.} \cite{Li2014A}, which is based on the extensive structural study on the ${\rm H_{2}S}$ compound at the pressure range of $\sim 10$~GPa - $200$~GPa. In the paper \cite{Li2014A} the following sequence of structural transitions was demonstrated: at the pressure of $8.7$~GPa the {\it Pbcm} structure is transformed into the {\it P2/c} structure. The next transformation occurs for the pressure of $29$~GPa ({\it P2/c}$\rightarrow${\it Pc}). For the pressure of $65$~GPa the transformation of {\it Pc} structure into the {\it $Pmc2_{1}$} structure was observed. It is worth paying attention to the fact that the obtained theoretical results are consistent with 
the X-ray diffraction (XRD) experimental data \cite{Endo1996A, Fujihisa1998A}. The last two structural transitions for compressed ${\rm H_{2}S}$ were observed for $80$~GPa ({\it $Pmc2_{1}$}$\rightarrow${\it P-1}) and $160$~GPa ({\it P-1}$\rightarrow${\it Cmca}). Interestingly, the results 
obtained in the paper \cite{Li2014A} are in the contradiction with the earlier theoretical predictions, suggesting that ${\rm H_{2}S}$ dissociate into elemental sulfur and hydrogen under the high pressure \cite{Rousseau2000A}. However, it should be noted, that the partial decomposition of ${\rm H_{2}S}$ was observed in Raman \cite{Sakashita2000A} and XRD studies \cite{Fujihisa2004A} at the room temperature above $27$~GPa. The calculations of the electronic structure made in \cite{Li2014A} suggest that the ${\rm H_{2}S}$ compound is the insulator up to the pressure of $130$~GPa. This result correlates well with the value of the metallization pressure of about $96$~GPa, observed experimentally \cite{Sakashita1997A} 
(see also related research \cite{Shimizu1991A, Endo1994A, Endo1996A, Endo1998A, Fujihisa1998A, Fujihisa2004A, Shimizu1992A, Shimizu1995A, Shimizu1997A, Loveday2000A, Sakashita2000A}). It is worth noting that in the work \cite{Li2014A} for the structure of {\it Pbcm } ($p=0.3$~GPa) the large indirect band gap of $\sim5.5$~eV was determined, which is relatively good comparing with the experimental results ($4.8$~eV) \cite{Kume2002A}. With the pressure increase, the band gap decreases ($3.75$~eV for $15$~GPa, $1.6$~eV for $40$~GPa, and $0.27$~eV for $120$~GPa). 
For the metallization pressure ($130$~GPa), the value of the electron density of states ($\rho\left(\varepsilon_{F}\right)$) is $0.33$~$\rm eV^{-1}$ per f.u. The sudden increase in $\rho\left(\varepsilon_{F}\right)$ to the value of $0.51$~$\rm eV^{-1}$ per f.u. is observed at the structural transition {\it P-1}$\rightarrow${\it Cmca} \cite{Li2014A}.  

In experiments described in the papers \cite{Drozdov2014A} and \cite{Drozdov2015A} are observed two different superconducting states. 
In particular, superconductivity measured in the low-temperature range (l-T, sample prepared at $T<100$~K), possibly relates to the ${\rm H_{2}S}$ compound, as it is generally consistent with calculations presented in \cite{Li2014A} for solid ${\rm H_{2}S}$: both the value of $T_{C}<82$~K and its pressure behavior. In addition, it was demonstrated theoretically \cite{Durajski2015A} that the experimental results could be reproduced accurately in the framework of classical Eliashberg equations, whereas the value of Coulomb pseudopotential is low ($\mu^{\star}\sim 0.15$).

On the other hand, the result obtained by Ishikawa {\it et al.} \cite{Ishikawa2016A} suggest that in the narrow pressure range from $110$~GPa to $123$~GPa, the ${\rm H_{5}S_{2}}$ compound, in which asymmetric hydrogen bonds are formed between ${\rm H_{2}S}$ and ${\rm H_{3}S}$ molecules, is thermodynamically stable and its critical temperature correlates well with experimental results \cite{Drozdov2014A, Drozdov2015A}. 
However, it should be assumed that the anharmonic effects lower the theoretically determined value of the critical temperature by the minimum of about 
$20$\%. This assumption has some theoretical justification \cite{Errea2015A, Errea2016A}, nevertheless in the paper \cite{Ishikawa2016A}, it is not supported by the first-principles calculations.

It is also worth paying attention to the results contained in \cite{Li2016A}, in which it is envisaged that the ${\rm H_{4}S_{3}}$ is stable within the pressure range of $25$-$113$~GPa. What is important, ${\rm H_{4}S_{3}}$ coexists with fraction of ${\rm H_{3}S}$ and ${\rm H_{2}S}$, at least up to the pressure of $140$~GPa. The theoretical results correlate with XRD data, which confirm that above $27$~GPa, the dihydrogen sulfide partially decomposes into ${\rm S+H_{3}S + H_{4}S_{3}}$. Nevertheless, ${\rm H_{4}S_{3}}$ is characterized by the very low critical temperature 
($T_{C}=2.2$~K for $\mu^{\star}=0.13$), which suggests that kinetically protected ${\rm H_{2}S}$ in samples prepared at low temperature is responsible for the observed superconductivity below $160$~GPa \cite{Li2016A}.

The superconductivity with the record high value of critical temperature ($T_{C}=203$~K and $p=155$~GPa), 
obtained for the sample prepared at high temperatures (h-T), relates to the decomposition of starting material above $43$~GPa \cite{Duan2015A} 
(see also \cite{Bernstein2015A, Errea2015A, Li2016A, Einaga2016A}): ${\rm 3H_{2}S\rightarrow 2H_{3}S + S}$. 
We notice that the experimental values of critical temperature and its pressure dependencies are close to the values of $T_{C}$ predicted theoretically 
by  Duan {\it et al.} \cite{Duan2014A, Duan2015A} ($T_{C}\in\left<191-204\right>$~K at $200$~GPa), or later by Errea {\it et al.} \cite{Errea2016A} for the cubic {\it Im-3m} structure with ${\rm H_{3}S}$ stoichiometry. The physical mechanism underlying the superconductivity of ${\rm H_{3}S}$ is similar to that in the ${\rm MgB_{2}}$ compound: metallization of covalent bonds. The main difference from the magnesium dibore is that the hydrogen mass is $11$ times smaller than the mass of boron \cite{Bernstein2015A}. Considering the electron-phonon interaction, it was noted that in the case of ${\rm H_{3}S}$, 
the vertex corrections in the local approximation have the least meaning \cite{Durajski2016A}. However, in the static limit with finite {\bf q}, 
the vertex corrections change the critical temperature by $-34$~K \cite{Sano2016A}. The result above correlates well with the data for ${\rm SiH_{4}}$ \cite{Wei2010A}. For the ${\rm H_{3}S}$ compound, the thermodynamic parameters are also affected by the anharmonic effects. It was shown that for $200$~GPa and $250$~GPa, the anharmonic effects lower substantially the value of electron-phonon coupling constant 
\cite{Errea2015A, Sano2016A}. It is worth noting that the pressure which increases above $250$~GPa does not increase the critical temperature value 
in ${\rm H_{3}S}$ \cite{Durajski2017A}. However, we have recently shown that the value of $T_{C}$ increases ($242$~K), if we use the sulfur isotope 
${\rm ^{36}S}$ \cite{Szczesniak2018A}. Therefore, it can be reasonably supposed that it is possible to obtain the superconducting condensate close to the room temperature (see also related research \cite{Peng2017A}).

In the presented paper, we analyze precisely the thermodynamic properties of superconducting state in the ${\rm H_{5}S_{2}}$ system \cite{Ishikawa2016A}. According to what we have mentioned before this compound is thermodynamically stable in the narrow pressure range from $110$~GPa to $123$~GPa. As the part of the analysis, we will prove that for the pressure at $112$~GPa, the superconducting state is characterized by the anomalously high value of $\mu^{\star}$ (also after taking into account the vertex corrections to the electron-phonon interaction). The above result is not consistent with the result obtained by Ishikawa {\it et al.} \cite{Ishikawa2016A}, where $\mu^{\star}\in\left<0.13,0.17\right>$. 
The paper contains also the parameter's analysis of superconducting state, that is induced in the ${\rm H_{4}S_{3}}$ compound \cite{Li2016A}.   
The results obtained by us were compared with the results for the ${\rm H_{2}S}$ compound, which the thermodynamic properties 
of superconducting state naturally explain the properties of low-temperature superconducting phase in the compressed hydrogen sulfide \cite{Li2014A, Durajski2015A}.

\section{FORMALISM}

Let us take into account the Eliashberg equations on the imaginary axis ($i=\sqrt{-1}$):
\begin{widetext}
\begin{eqnarray}
\label{r1-II}
\varphi_{n}&=&\pi k_{B}T\sum_{m=-M}^{M}
\frac{\lambda_{n,m}-\mu_{m}^{\star}}
{\sqrt{\omega_m^2Z^{2}_{m}+\varphi^{2}_{m}}}\varphi_{m}\\ \nonumber
&-&
A\frac{\pi^{3}\left(k_{B}T\right)^{2}}{4\varepsilon_{F}}\sum_{m=-M}^{M}\sum_{m'=-M}^{M}
\frac{\lambda_{n,m}\lambda_{n,m'}}
{\sqrt{\left(\omega_m^2Z^{2}_{m}+\varphi^{2}_{m}\right)
       \left(\omega_{m'}^2Z^{2}_{m'}+\varphi^{2}_{m'}\right)
       \left(\omega_{-n+m+m'}^2Z^{2}_{-n+m+m'}+\varphi^{2}_{-n+m+m'}\right)}}\\ \nonumber
&\times&
\left[
\varphi_{m}\varphi_{m'}\varphi_{-n+m+m'}+2\varphi_{m}\omega_{m'}Z_{m'}\omega_{-n+m+m'}Z_{-n+m+m'}-\omega_{m}Z_{m}\omega_{m'}Z_{m'}
\varphi_{-n+m+m'}
\right],
\end{eqnarray}
and
\begin{eqnarray}
\label{r2-II}
Z_{n}&=&1+\frac{\pi k_{B}T}{\omega_{n}}\sum_{m=-M}^{M}
\frac{\lambda_{n,m}}{\sqrt{\omega_m^2Z^{2}_{m}+\varphi^{2}_{m}}}\omega_{m}Z_{m}\\ \nonumber
&-&
A\frac{\pi^{3}\left(k_{B}T\right)^{2}}{4\varepsilon_{F}\omega_{n}}\sum_{m=-M}^{M}\sum_{m'=-M}^{M}
\frac{\lambda_{n,m}\lambda_{n,m'}}
{\sqrt{\left(\omega_m^2Z^{2}_{m}+\varphi^{2}_{m}\right)
       \left(\omega_{m'}^2Z^{2}_{m'}+\varphi^{2}_{m'}\right)
       \left(\omega_{-n+m+m'}^2Z^{2}_{-n+m+m'}+\varphi^{2}_{-n+m+m'}\right)}}\\ \nonumber
&\times&
\left[
\omega_{m}Z_{m}\omega_{m'}Z_{m'}\omega_{-n+m+m'}Z_{-n+m+m'}+2\omega_{m}Z_{m}\varphi_{m'}\varphi_{-n+m+m'}-\varphi_{m}\varphi_{m'}\omega_{-n+m+m'}Z_{-n+m+m'}
\right].
\end{eqnarray}
\end{widetext}
In the case when $A=1$, the Eliashberg set was generalized to include the lowest-order vertex correction - scheme VCEE 
({\bf V}ertex {\bf C}orrected {\bf E}liashberg {\bf E}quations) \cite{Freericks1998A}. On the other hand ($A=0$), we get the model without the vertex corrections: the so-called CEE scheme ({\bf C}lassical {\bf E}liashberg {\bf E}quations) \cite{Eliashberg1960A}. Note that in the considered equations, 
the momentum dependence of electron-phonon matrix elements has been neglected, which is equivalent to using the local approximation.

The individual symbols in equations \eq{r1-II}-\eq{r2-II} have the following meaning:
$\varphi_{n}=\varphi\left(i\omega_{n}\right)$ is the order parameter function and $Z_{n}= Z\left(i\omega_{n}\right)$ represents the wave function renormalization factor. The quantity $Z_{n}$ describes the renormalization of the thermodynamic parameters of superconducting state by the electron-phonon interaction \cite{Carbotte1990A}. This is the typical strong-coupling effect, because $Z_{n=1}$ in the CEE scheme is expressed by the formula: $Z_{n=1}\sim 1+\lambda$, where $\lambda$ denotes the electron-phonon coupling constant: 
$\lambda=2\int^{\omega_{D}}_{0}d\omega\alpha^{2}F\left(\omega\right)/\omega$, and $\omega_{D}$ is the Debye frequency. The order parameter is defined as 
the ratio: $\Delta_{n}=\phi_{n}/Z_{n}$. Symbol $\omega_{n}$ is the Matsubara frequency: $\omega_{n}=\pi k_{B}T\left(2n+1\right)$, while $n$ is the integer. Let us emphasize that the dependence of order parameter on the Matsubara frequency means that the Eliashberg formalism explicitly takes into account the retarding nature of electron-phonon interaction. In the present paper, it was assumed $M=1100$, which allowed to achieve convergent results in the range from $T_{0}=5$~K do $T_{C}$ (see also \cite{Szczesniak2006A}).

The function $\mu^{\star}_{n}$ modeling the depairing correlations has the following form: 
$\mu_{n}^{\star}=\mu^{\star}\theta \left(\omega_{c}-|\omega_{n}|\right)$. The Heaviside function is given by $\theta\left(x\right)$ and $\omega_{c}$ represents the cut-off frequency. By default, it is assumed that $\omega_{c}\in\left<3\omega_{D}, 10\omega_{D}\right>$.

The electron-phonon pairing kernel is expressed by the formula:
\begin{equation}
\label{r3-II}
\lambda_{n,m}=2\int_0^{\omega_{D}}d\omega\frac{\omega}{\omega ^2+4\pi^{2}\left(k_{B}T\right)^{2}\left(n-m\right)^{2}}\alpha^{2}F\left(\omega\right).
\end{equation}
The higher order corrections are not included in equation \eq{r3-II}. The full formalism demands taking into consideration the additional terms related to the phonon-phonon interactions and the non-linear coupling between the electrons and the phonons. This has been discussed by Kresin {\it et al.} in the paper \cite{Kresin1993A}.
\begin{figure*}
\includegraphics[width=\columnwidth]{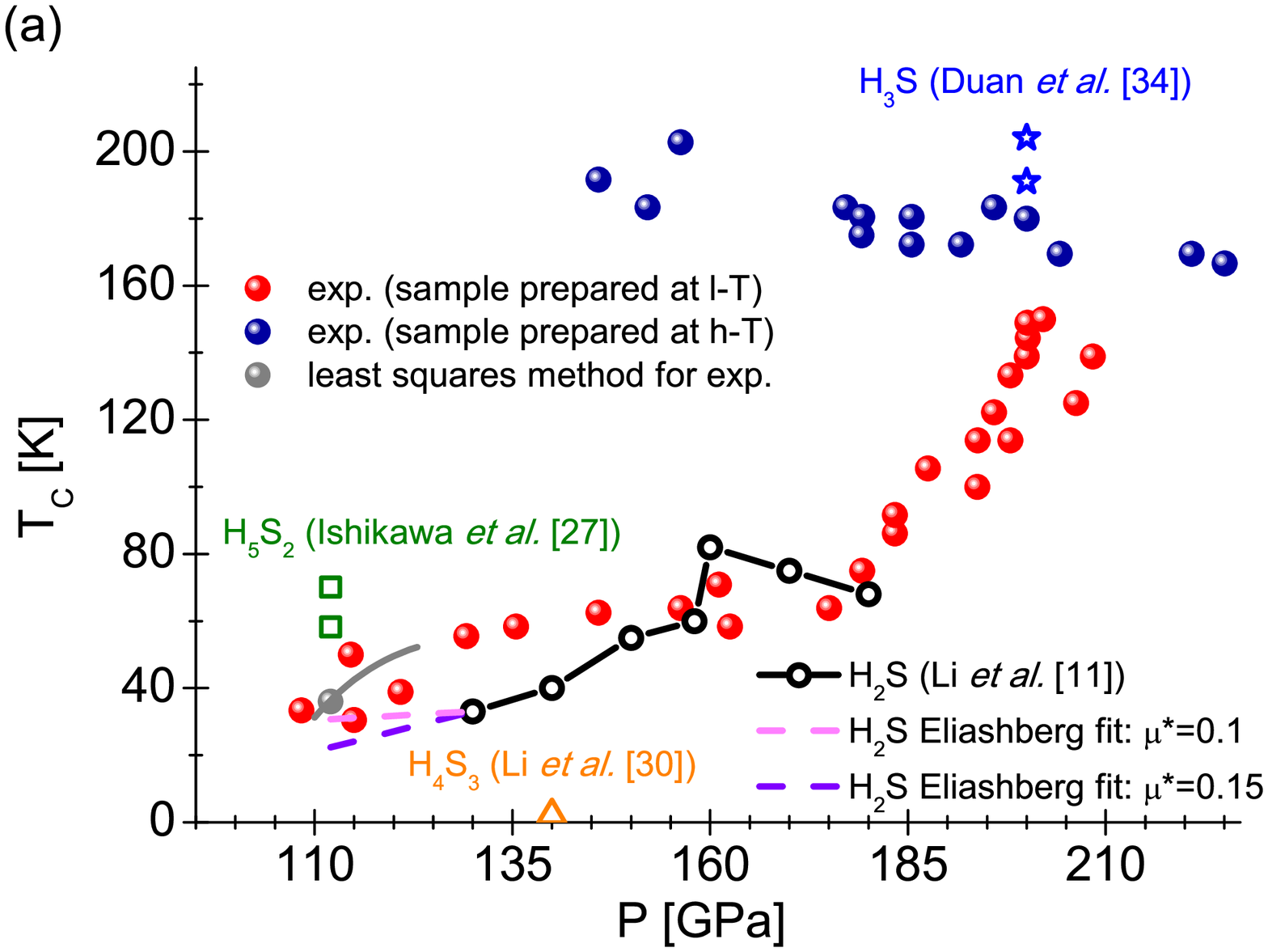}
\includegraphics[width=\columnwidth]{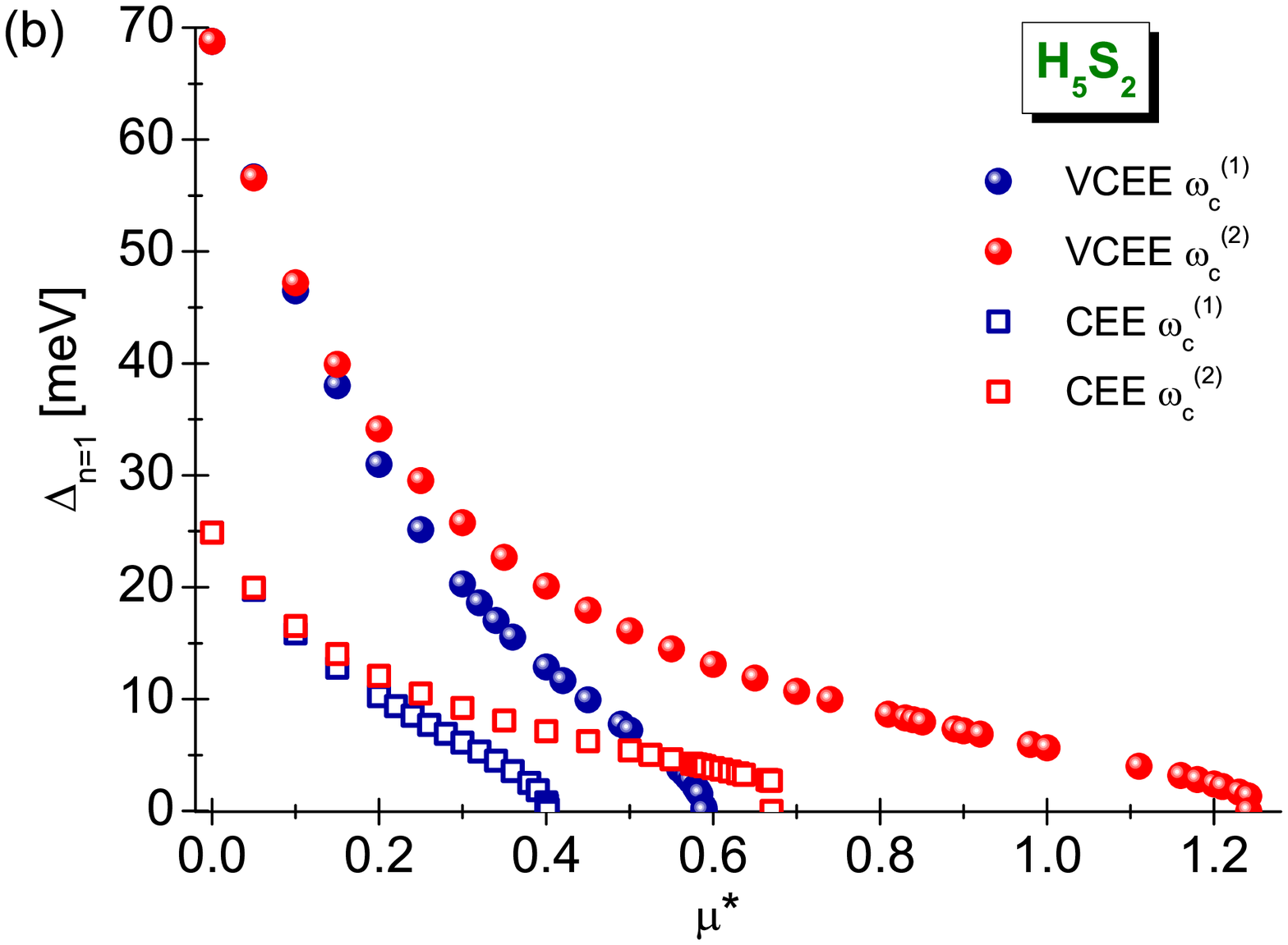}
\caption{(a) The experimental values of critical temperature as a function of pressure for the selected hydrogen-containing compounds of sulfur. 
             The following designation was concluded in the legend: exp. - experimental data taken from the paper \cite{Drozdov2015A}.
             The gray line represents the dependence of $T_{C}\left(p\right)$ in the pressure range from  $110$~GPa to $123$~GPa, for which 
             Ishikawa {\it et al.} \cite{Ishikawa2016A} predicted the stability of  ${\rm H_{5}S_{2}}$ compound. 
             The black line with the circles indicates the theoretical results, obtained for ${\rm H_{2}S}$ \cite{Li2014A}. 
             The violet and magenta dash lines are the predictions at the phenomenological level based on the classical theory of Eliashberg.  
             In addition, we have included theoretical results for ${\rm H_{5}S_{2}}$ \cite{Ishikawa2016A} (green squares), 
             $\rm H_{4}S_{3}$ \cite{Li2016A} (orange triangle) and 
             $\rm H_{3}S$ (blue stars) \cite{Duan2014A}. 
         (b) The dependence of order parameter ($\Delta_{n=1}$) on the Coulomb pseudopotential for ${\rm H_{5}S_{2}}$  -
             selected values of cut-off frequency ($T=T_{C}$)}. 
             The spheres represent the results obtained with the aid of Eliashberg equations with the vertex corrections (VCEE), 
             the squares correspond to the results obtained in the framework of Migdal-Eliashberg formalism (CEE).
\label{f01}
\end{figure*}
The Eliashberg spectral function is defined as:
\begin{equation}
\label{r4-II}
\alpha^2F(\omega) = {1\over 2\pi \rho\left(\varepsilon_{F}\right)}\sum_{{\bf q}\nu} 
                    \delta(\omega-\omega_{{\bf q}\nu})
                    {\gamma_{{\bf q}\nu}\over\omega_{{\bf q}\nu}},
\end{equation}
with:
\begin{eqnarray}
\label{r5-II}
\gamma_{{\bf q}\nu}&=&2\pi\omega_{{\bf q}\nu} \sum_{ij}
                \int {d^3k\over \Omega_{BZ}}  |g_{{\bf q}\nu}({\bf k},i,j)|^2\\ \nonumber
                    &\times&\delta(\varepsilon_{{\bf q},i}-\varepsilon_{F})\delta(\varepsilon_{{\bf k}+{\bf q},j}-\varepsilon_{F}), 
\end{eqnarray}
where $\omega_{{\bf q}\nu}$ determines the values of phonon energies and $\gamma_{{\bf q}\nu}$ represents the phonon linewidth. The electron-phonon coefficients are given by $g_{{\bf q}\nu}({\bf k},i,j)$ and $\varepsilon_{{\bf k},i}$ is the electron band energy ($\varepsilon_{F}$ denotes the Fermi energy). 

For the purpose of this paper, the spectral functions obtained by: 
 Li {\it et al.} \cite{Li2014A} (${\rm H_{2}S}$), Ishikawa {\it et al.} \cite{Ishikawa2016A} (${\rm H_{5}S_{2}}$), 
 and Li {\it et al.} \cite{Li2016A} (${\rm H_{4}S_{3}}$), have been taken into account. The Eliashberg functions were calculated in the framework of Quantum Espresso code \cite{Baroni1986A, Giannozzi2009A}.  
 
\section{RESULTS}
%

\subsection{The pseudopotential parameter}
\begin{figure}
\includegraphics[width=\columnwidth]{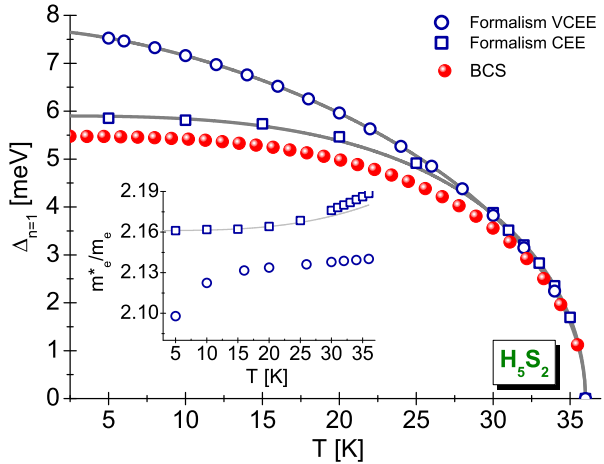}
\caption{The influence of temperature on the maximum value of order parameter in the VCEE and CEE scheme for ${\rm H_{5}S_{2}}$ 
         ($\omega_{c}=\omega^{\left(1\right)}_{c}$). 
         The insertion presents the ratio $m^{\star}_{e} /m_{e}$ as a function of temperature.  
         The open symbols represent the numerical results, the gray lines correspond to the results obtained with the help of analytical formulas 
         (\eq{r1-IIIB} or \eq{r2-IIIB}). The red spheres were obtained in the BCS scheme assuming that:
         $2\Delta_{n=1}\left(0\right)/k_{B}T_{C}=3.53$ \cite{Bardeen1957A, Bardeen1957B}.}
\label{f02}
\end{figure}
\begin{figure}
\includegraphics[width=\columnwidth]{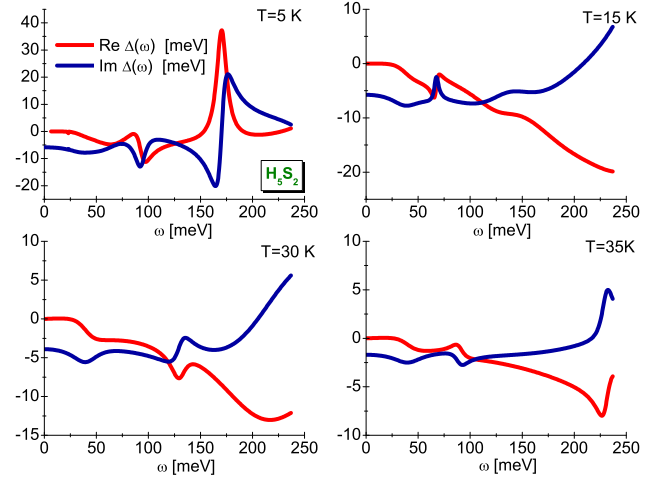}
\caption{The real and imaginary part of ${\rm H_{5}S_{2}}$ order parameter on the real axis for the selected values of temperature. 
         The sample results obtained in the CEE scheme.}
\label{f03}
\end{figure}

\fig{f01}~(a) presents the experimental dependence of critical temperature on the pressure for the sulfur-hydrogen systems 
\cite{Drozdov2015A} (see also \cite{Einaga2016A}). Additionally, we included the theoretical results obtained for the 
${\rm H_{5}S_{2}}$ \cite{Ishikawa2016A}, ${\rm H_{2}S}$ \cite{Li2014A},  $\rm H_{4}S_{3}$ \cite{Li2016A}, and 
$\rm H_{3}S$ \cite{Duan2014A}.

It is worth to notice that ${\rm H_{5}S_{2}}$ is thermodynamically stable in the fairly narrow pressure range ($110$~GPa - $123$~GPa). Especially at pressure $110$~GPa there is the transformation: ${\rm H_{4}S_{3}}+7{\rm H_{3}S}\rightarrow 5{\rm H_{5}S_{2}}$. On the other hand, breakup ${\rm H_{5}S_{2}}$ for $p=123$~GPa takes place according to the scheme: $3{\rm H_{5}S_{2}}\rightarrow 5{\rm H_{3}S}+\rm{S}$. 

The measuring points were used to determine the curve $T_{C}\left(p\right)$. We used the approximating function, matching the polynomial with the least squares method, respectively. On this basis, the estimated value of critical temperature for the pressure at $112$~GPa is equal to $36$~K. 
In the case of ${\rm H_{5}S_{2}}$ compound with value of $T_{C}=36$~K, we have calculated the pseudopotential parameter. To this end, we used the equation: 
$\left[\Delta_{n=1}\left(\mu^{\star}\right)\right]_{T=T_{C}}=0$ \cite{Szczesniak2013A}. We obtained the very high value of $\mu^{\star}$ 
in both considered approaches: $\left[\mu^{\star}\left(\omega^{\left(1\right)}_{c}\right)\right]_{VCEE}=0.589$ and 
$\left[\mu^{\star}\left(\omega^{\left(1\right)}_{c}\right)\right]_{CEE}=0.402$, whereby we have chosen the following cut-off frequency: 
$\omega^{\left(1\right)}_{c}=3\omega_{D}$, where $\omega_{D}=237.2$~meV \cite{Ishikawa2016A}. Additionally for ${\rm H_{5}S_{2}}$, we have 
$\varepsilon_{F}=22.85$~eV.

Note that the high value of $\mu^{\star}$ is relatively often observed in the case of high-pressure superconducting state. 
For example, for phosphorus: $\left[\mu^{\star}\left(5\omega_{D}\right)\right]^{p=20{\rm GPa}}_{CEE}=0.37$, where $\omega_{D}=59.4$~meV \cite{Duda2016A}. We encounter the similar situation for lithium: $\left[\mu^{\star}\left(3\omega_{D}\right)\right]^{p=29.7{\rm GPa}}_{CEE}=0.36$, while 
$\omega_{D}=82.7$~meV \cite{Szczesniak2010A}. Interestingly, for $\rm H_{3}S$ compound under the pressure at $150$~GPa, the value of Coulomb pseudopotential is low: $0.123$ \cite{Durajski2016B}. However, after increasing the pressure by $50$~GPa, this value is clearly increasing: 
$\mu^{\star}\sim 0.2$ \cite{Durajski2016A}. In the case of another compound ($\rm PH_{3}$), for which we also know the experimental critical temperature ($T_{C}=81$~K), the much lower value of Coulomb pseudopotential was obtained: $\left[\mu^{\star}\left(10\omega_{D}\right)\right]^{p=200{\rm GPa}}_{CEE}=0.088$ \cite{Durajski2016A}. It is also worth mentioning the paper \cite{Szczesniak2013A}, where the properties of superconducting state were studied in the $\rm SiH_{4}$ compound for $\mu^{\star}\in<0.1,0.3>$. The result was the decreasing critical temperature range from $51.7$~K to $20.6$~K, what in relation to the results contained in the experimental work \cite{Eremets2008A} suggests $\mu^{\star}\sim 0.3$. However, it should be remembered that the results presented in the publication \cite{Eremets2008A} are very much undermined \cite{Degtyareva2009A}, while it is argued that the experimental data does not refer to $\rm SiH_{4}$ but to the $\rm PtH$ compound (the hydrogenated electrodes of measuring system).

Considering the facts presented above, it is easy to see that the $\rm H_{5}S_2$ compound, even in the group of high-pressure superconductors, has the unusually high value of Coulomb pseudopotential. This is the feature of $\rm H_{5}S_2$ system, because values of $\mu^{\star}$ cannot be reduced by selecting another acceptable cut-off frequency. On the contrary, increasing $\omega_{c}$ leads to the absurdly large increase in the value 
of Coulomb pseudopotential: $\left[\mu^{\star}\left(\omega^{\left(2\right)}_{c}\right)\right]_{VCEE}=1.241$ and
$\left[\mu^{\star}\left(\omega^{\left(2\right)}_{c}\right)\right]_{CEE}=0.671$, where $\omega^{\left(2\right)}_{c}=10\omega_{D}$. 
The dependence courses of $\Delta_{n=1}$ on $\mu^{\star}$, characterizing the situation discussed by us, are collected in \fig{f01}~(b).

It should be emphasized that value of $\mu^{\star}$ calculated by us for the CEE case ($0.402$) significantly exceeds the value of Coulomb pseudopotential estimated in the paper \cite{Ishikawa2016A}, where $\mu^{\star}\in\left<0.13,0.17\right>$. This result is related to the fact that in the publication \cite{Ishikawa2016A} the critical temperature was calculated using the simplified Allen-Dynes formula ($f_{1}=f_{2}=1$) \cite{Allen1975A}, which significantly understates $\mu^{\star}$, for values greater than $0.1$ \cite{Szczesniak2013B}. 
This is due to the approximations used in the derivation of Allen-Dynes formula. Among other things, the analytical approach does not take into account the effects of retardation and the impact on the result of the cut-off frequency. Below is the explicit form of Allen-Dynes formula:
\begin{equation}
\label{r1-IIIA}
k_{B}T_{C}=f_{1}f_{2}\frac{\omega_{\ln}}{1.2}\exp\left[\frac{-1.04\left(1+\lambda\right)}{\lambda-\mu^{\star}\left(1+0.62\lambda\right)}\right],
\end{equation}
where:
\begin{equation}
\label{r2-IIIA}
f_{1}=\left[1+\left(\frac{\lambda}{\Lambda_{1}}\right)^{\frac{3}{2}}\right]^{\frac{1}{3}},
\end{equation}
\begin{equation}
\label{r3-IIIA}
\qquad f_{2}=1+\frac{\left(\frac{\sqrt{\omega_{2}}}{\omega_{\rm{ln}}}-1\right)\lambda^{2}}{\lambda^{2}+\Lambda^{2}_{2}}.
\end{equation}
Parameters $\Lambda_{1}$ and $\Lambda_{2}$ were calculated using the formulas:
\begin{equation}
\label{r4-IIIA}
\Lambda_{1}=2.46\left(1+3.8\mu^{\star}\right),
\end{equation}
\begin{equation}
\label{r5-IIIA}
\Lambda_{2}=1.82\left(1+6.3\mu^{\star}\right)\frac{\sqrt{\omega_{2}}}{\omega_{\ln}}.
\end{equation}
The second moment is given by the expression:
\begin{equation}
\label{r6-IIIA}
\omega_{2}=\frac{2}{\lambda}\int_{0}^{\omega_{D}}d\omega\alpha^{2}F\left(\omega\right)\omega.
\end{equation}
The quantity $\omega_{\ln}$ stands for the phonon logarithmic frequency:
\begin{equation}
\label{r7-IIIA}
\omega_{\ln}=\exp\left[\frac{2}{\lambda}\int^{\omega_{D}}_{0}d\omega\frac{\alpha^{2}F\left(\omega\right)}{\omega}\ln\left(\omega\right)\right].
\end{equation}

For the $\rm{H_5S_2}$ compound, we have obtained: $\omega_{2}=112.88$~meV and $\omega_{\ln}=77.37$~meV. Assuming $f_{1}=f_{2}=1$, we reproduce the result contained in \cite{Ishikawa2016A}: 
$\left[\rm{T_{C}}\right]^{\mu^{\star}=0.17}=58.3$~K and $\left[\rm{T_{C}}\right]^{\mu^{\star}=0.13}=70.1$~K. However, the full Allen-Dynes formula gives: 
$\left[\rm{T_{C}}\right]^{\mu^{\star}=0.17}=62.5$~K and $\left[\rm{T_{C}}\right]^{\mu^{\star}=0.13}=76.1$~K.

From the physical point of view, the values of Coulomb pseudopotential calculated for ${\rm H_{5}S_{2}}$, in the framework of full Eliashberg formalism, are so high that $\mu^{\star}$ cannot be associated only with the depairing Coulomb correlations. In principle, it should be treated only as the parameter to fit the model to the experimental data. Of course, there can be very much reasons that cause anomalously high value of $\mu^{\star}$. 
The authors of paper \cite{Ishikawa2016A} pay special attention to the role of anharmonic effects. It should be emphasized that even if it were this way, the anharmonic nature of phonons should be taken into account not only in the Eliashberg function, but also in the structure of Eliashberg equations itself (this fact is usually omitted in the analysis scheme). In our opinion, the probable cause may also be non-inclusion in the Fr{\"o}hlich Hamiltonian \cite{Frohlich1952A} the non-linear terms of electron-phonon interaction \cite{Kresin1993A} or possibly anomalous electron-phonon interactions related to the dependence of full electron Hamiltonian parameters of system on the distance between the atoms of crystal lattice 
\cite{Szczesniak2012A, Szczesniak2017A}. 

However, in our opinion it is more likely that the experimental results obtained for the low temperature superconducting state of compressed sulfur-hydrogen system are largely related to the condensate being induced in the ${\rm H_{2}S}$ compound (see black line in \fig{f01}). 
In fact, the predicted thermodynamically stable pressure range of ${\rm H_{5}S_{2}}$ is really narrow ($110$-$123$~GPa), 
and the inevitable kinetic barrier may prevent the decomposition of ${\rm H_{2}S}$ into ${\rm H_{5}S_{2}}$ in such narrow pressure range. 
It should also be mentioned that in the following two experiments \cite{Li2016A} and \cite{Einaga2016A}, the XRD measurements on compressed ${\rm H_{2}S}$ do not observe the formation of ${\rm H_{5}S_{2}}$. On the other hand the XRD results in the paper \cite{Li2016A} indeed observed the residual 
${\rm H_{2}S}$ coexist with dissociation products ${\rm H_{3}S}$ and ${\rm H_{4}S_{3}}$ at least up to $140$~GPa. It should also be emphasized that the anomalously high $\mu^{\star}$ obtained for ${\rm H_{5}S_{2}}$ are extremely incompatible with the estimation of $\mu^{\star}$ value ($0.1$-$0.13$) contained in Ashcroft's fundamental paper \cite{Ashcroft2004A} concerning the superconducting state in the hydrogen-rich compounds.

The value of Coulomb pseudopotential observed in ${\rm H_{5}S_{2}}$ compound is so high that it is worth to confront it with other microscopic parameters obtained from {\it ab initio} calculations: ($\varepsilon_{F}$ and $\omega_{{\rm \ln}}$) \cite{Ishikawa2016A}. The most advanced formula on the Coulomb pseudopotential takes the following form \cite{Bauer2012A}:
\begin{equation}
\label{r8-IIIA}
\mu^{\star}=\frac{\mu+a\mu^2}{1+\mu \ln{(\frac{\varepsilon_{F}}{\omega_{{\rm \ln}}})}+a\mu^2\ln{(\frac{\alpha \varepsilon_{F}}{\omega_{{\rm \ln}}})}},
\end{equation}
where $a=1.38$ and $\alpha\simeq 0.1$. Symbol $\mu$ denotes product of the electron density of states at the Fermi level 
$\rho\left(\varepsilon_{F}\right)$ and the Coulomb potential $U>0$. Note that the expression \eq{r8-IIIA} is the non-trivial generalization 
of Morel-Anderson formula \cite{Morel1962A}:
\begin{equation}
\label{r9-IIIA}
\mu^{\star}=\frac{\mu}{1+\mu \ln{(\frac{\varepsilon_{F}}{\omega_{{\rm \ln}}})}}.
\end{equation}
In particular, with the help of equation \eq{r8-IIIA}, it can be proven that the retardation effects associated with the electron-phonon interaction reduce the original value of $\mu$ to the value of $\mu^{\star}$, but to the much lesser extent than expected by Morel and Anderson. 

Using the formula \eq{r8-IIIA}, it is easy to show that the anomalously high value of $\mu^{\star}$ cannot result only from the strong electron depairing correlations. Namely in the limit $\mu\rightarrow +\infty$, we obtain:
$\left[\mu^{\star}\right]_{\rm max}=1/\ln\left(\alpha\varepsilon_{F}/\omega_{{\rm ln}}\right)=0.295$. 
It is worth noting that the Morel-Anderson model is completely unsuitable for the analysis, because for $\mu^{\star}=0.402$ we get the negative (non-physical) value of $\mu=-0.48$.

The value of Coulomb pseudopotential can also be tried to be estimated using the phenomenological Bennemann-Garland formula \cite{Bennemann1972A}:
\begin{equation}
\label{r10-IIIA}
\mu^{\star}_{\rm BG}\sim 0.26 \rho\left(\varepsilon_{F}\right)/[1+\rho\left(\varepsilon_{F}\right)].
\end{equation}

For the $\rm{H_5S_2}$ compound subjected to the action of pressure at $112$~GPa, we obtain $\mu^{\star}_{\rm BG} = 0.23$, which also proves the breakdown 
of the classical interpretation of $\mu^{\star}$.

Taking into account all the facts given above, it is unlikely for the low-temperature phase of compressed sulfur-hydrogen to be tied with  ${\rm H_{5}S_{2}}$. The dependence of critical temperature on the pressure could be explained according to the ${\rm H_{2}S}$ compound, whereas the low   value of Coulomb pseudopotential is assumed ($\mu^{\star}=0.15$) \cite{Durajski2015A, Li2014A}. In particular, let's focus on the lowest pressure considered in \cite{Li2014A} ($130$~GPa). The calculations within the paper, are  performed according to the CEE scheme and they give 
the critical temperature value of $\sim 30.6$~K \cite{Durajski2015A}, which correlates well with the experimental value of $T_{C}=36$~K for $p=112$~GPa. 
It is worth noting that the electron-phonon interaction in ${\rm H_{2}S}$ for the pressure of $130$~GPa, is characterized by the following set of parameters:  $\lambda=0.785$, $\omega_{\rm ln}=81.921$~meV, and $\sqrt{\omega_{2}}=112.507$~meV. 
Physically, this means the intermediate value of coupling between the electrons and phonons. ${\rm H_{5}S_{2}}$ is the system characterized by the strong electron-phonon coupling  ($\lambda=1.186$). At the phenomenological level, one can even extend the  $T_{C}\left(p\right)$ curve calculated in the paper \cite{Li2014A}. For this purpose, with the help of the approximation function, we determined the relationship $\lambda\left(p\right)$, which we used in the appropriately modified Eliashberg equations (the case of one-band equations presented in \cite{Szczesniak2013C}). As the result for $p=112$~GPa, we received $T_{C}\in\left<30.7,22.3\right>$~K assuming that $\mu^{\star}\in\left<0.1,0.15\right>$. The full form of $T_{C}\left(p\right)$ curves, 
for $\mu^{\star}=0.1$ and $\mu^{\star}=0.15$ in the pressure range from $112$~GPa to $130$~GPa, is presented in \fig{f01}~(a).

It is also worth mentioning the superconducting state, which can potentially be induced in ${\rm H_{4}S_{3}}$. 
The data obtained for the pressure of $140$~GPa gives the following characteristics of the electron-phonon interaction \cite{Li2016A}: $\lambda=0.42$ and 
$\omega_{{\rm \ln}}= 71.869$~meV (typical limit of weak coupling). Due to the low value of $\lambda$, the critical temperature can be calculated using the McMillan formula \cite{McMillan1968A} ($f_{1}=f_{2}=1$). Assuming that $\mu^{\star}=0.13$, we get $T_{C}=2.2$~K. This result means that the superconducting state inducing in ${\rm H_{4}S_{3}}$ cannot be equated with the low-temperature superconducting phase of compressed sulfur-hydrogen system. We probably have the analogous situation here as for the ${\rm H_{5}S_{2}}$ compound, where kinetically protected ${\rm H_{2}S}$ in samples prepared at low temperature is responsible for the observed superconductivity below $160$~GPa \cite{Li2016A}.

In the following chapters, for illustrative purposes, we calculated the thermodynamic parameters for the 
${\rm H_{5}S_{2}}$ compound. Note, that even for anomalously high value of $\mu^{\star}$, the classical Eliashberg equations allow to set thermodynamic 
functions of superconducting state correctly, but at the phenomenological level \cite{Carbotte1990A, Wiendlocha2016A}. The presented discussion is complemented by the results obtained in the CEE scheme for  ${\rm H_{2}S}$ ($p=130$~GPa) and $\rm H_{4}S_{3}$ ($p=140$~GPa) compounds.

\subsection{The order parameter and the electron effective mass}
\begin{figure}
\includegraphics[width=\columnwidth]{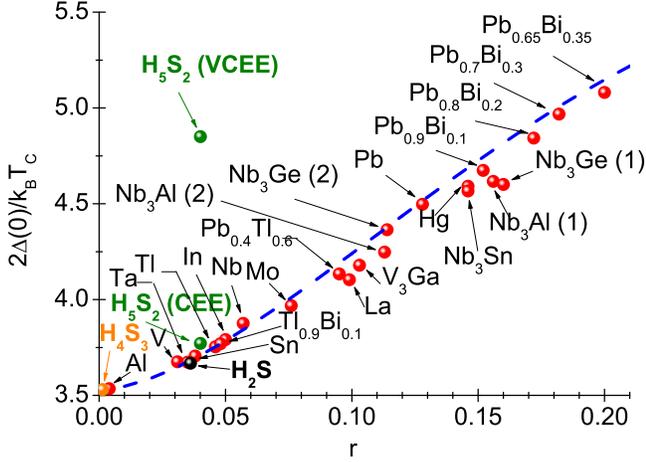}
\caption{The value of ratio $2\Delta\left(0\right)/k_{B}T_{C}$ in the dependence on the parameter $r=k_{B}T_{C}/\omega_{\ln}$. 
         The blue line represents the general trend reproduced by the formula:  
         $2\Delta\left(0\right)/k_{B}T_{C}=3.53 \left[1+12.5\left(r\right)^2\ln\left(1/2r\right)\right]$ (CEE scheme \cite{Carbotte1990A}). 
         }
\label{f04}
\end{figure}

\fig{f02} presents for ${\rm H_{5}S_{2}}$ the plot of the dependence of order parameter for the first Matsubara frequency on the temperature. It is clearly visible that, in the low temperature range, the values of order parameter calculated with regard to the vertex corrections are much higher than the values of $\Delta_{n=1}$ designated within the framework of classic Eliashberg scheme. This is the very interesting result, because the ratio 
$v_{s}=\lambda\omega_{D}/\varepsilon_{F}$ for ${\rm H_{5}S_{2}}$ compound is not high ($v_{s}=0.012$) \cite{Ishikawa2016A}. This result, with the cursory analysis, could suggest the slight influence of vertex corrections on the thermodynamic parameters. It is not, however, because it should be commemorated that $v_{s}$ defines only the static criterion of the impact of vertex corrections. This means that the differences observed between the results obtained in the VCEE and CEE schemes are associated with dynamic effects, {\it i.e.} the explicit dependence of order parameter on the Matsubara frequency. 

On the insertion in \fig{f02}, we presented the influence of temperature on the value of the ratio of electron effective mass 
($m^{\star}_{e}$) to the electron band mass ($m_{e}$). In the Eliashberg formalism this ratio can be estimated using the following formula: 
$m^{\star}_{e}/m_{e}=Z_{n=1}\left(T\right)$ \cite{Carbotte1990A}. It can be seen that, in the entire temperature range analyzed by us, the effective mass of electron is high – however slightly dependent on the temperature. In particular, in the VCEE scheme we get: 
$\left[m^{\star}_{e}\right]_{T_{0}}=2.10m_{e}$ and $\left[m^{\star}_{e}\right]_{T_{C}}=2.14m_e$.   
On the other hand, the classic Eliashberg approach gives:
$\left[m^{\star}_{e}\right]_{T_{0}}=2.16m_{e}$ and $\left[m^{\star}_{e}\right]_{T_{C}}=2.19m_e$. When comparing the above results, we conclude that the vertex corrections have the very little effect on the value of the effective mass of electron, contrary to the situation with the order parameter.

The functions plotted in \fig{f02} can be characterized analytically by means of formulas: 
\begin{equation}
\label{r1-IIIB}
\Delta_{n=1}\left(T\right)=\Delta_{n=1}\left(0\right)\sqrt{1-\left(T\slash T_{C}\right)^{\Gamma}}
\end{equation}
and (only for the CEE scheme): 
\begin{equation}
\label{r2-IIIB}
m_{e}^{\star}\slash m_{e}=\left[Z_{n=1}\left(T_{C}\right)-Z_{n=1}\left(0\right)\right]\left(T/T_{C}\right)^{\Gamma}
+Z_{n=1}\left(0\right),
\end{equation}
where the traditional markings were introduced: $\Delta_{n=1}\left(0\right)=\Delta_{n=1}\left(T_{0}\right)$ and 
$Z_{n=1}\left(0\right)=Z_{n=1}\left(T_{0}\right)$.

For the order parameter, we obtained the following estimation of temperature exponent:  
$\left[\Gamma\right]_{VCEE}=1.55$ and $\left[\Gamma\right]_{CEE}=3.15$. Physically, this result means that in the VCEE scheme the temperature dependence 
of order parameter differs very much from the course anticipated by the mean-field BCS theory, where $\Gamma_{\rm BCS}=3$ \cite{Eschrig2001A}. 
On the other hand, not very large deviations from the predictions of BCS theory at the level of classical Eliashberg equations can be explained 
by referring to the impact of retardation and strong-coupling effects on the superconducting state. In the simplest way, they are characterized by the ratio $r=k_{B}T_{C}/\omega_{\ln}$, which for the ${\rm H_{5}S_{2}}$ compound equals $0.0401$. On the other hand, in the BCS limit we obtain: $r=0$. 
Of course, in the case of the VCEE scheme, the deviations from the BCS predictions should be interpreted as the cumulative effect of vertex corrections, strong-coupling, and retardation effects influencing the superconducting state.

The very accurate values of order parameter can be calculated by referring to the equation:  
$\Delta\left(T\right)={\rm Re}\left[\Delta\left(\omega=\Delta\left(T\right)\right)\right]$, where the symbol $\Delta\left(\omega\right)$ denotes 
the order parameter determined on the real axis. The form of function $\Delta\left(\omega\right)$ should be determined using the method of the analytical extension of imaginary axis solutions \cite{Beach2000A}. The sample results have been posted in \fig{f03}, 
where it can be immediately noticed that the function of order parameter takes the complex values. However, for low values of frequency, 
only the real part of order parameter is non-zero. Physically, this means no damping effects, or equally, the endless life of Cooper pairs. 
Having the open form of order parameter on the real axis, we have calculated the value of ratio: $R_{\Delta}=2\Delta\left(0\right)/k_{B}T_{C}$. 
The obtained result was compared with the data for other superconductors with the phonon pairing mechanism (see \fig{f04}). 
It is easy to see that the result of CEE very well fits in the general trend anticipated by the classic Eliashberg formalism ($R_{\Delta}=3.77$). 
On the other hand, the impact of vertex corrections on the value of parameter $R_{\Delta}$ is strong ($R_{\Delta}=4.85$). 
Let us recall that the mean-field BCS theory predicts: $R_{\Delta}=3.53$ \cite{Bardeen1957A, Bardeen1957B}.

Particularly interesting is the comparison between the calculated values of $R_{\Delta}$ for ${\rm H_{5}S_{2}}$ with results obtained for ${\rm H_{2}S}$ and ${\rm H_{4}S_{3}}$. In the first step, it is noticeable that the value of parameter $r$ for ${\rm H_{2}S}$ and ${\rm H_{4}S_{3}}$ is  $0.0362$ and $0.0014$, respectively. It is evident that $\left[r\right]_{\rm H_{2}S}$ is close to the value of $r$ obtained for  ${\rm H_{5}S_{2}}$, while in the case of  ${\rm H_{4}S_{3}}$ the retardation and strong-coupling effects do not play the major role. 
In the CEE scheme, the dimensionless ratio $R_{\Delta}$ for ${\rm H_{2}S}$ and ${\rm H_{4}S_{3}}$ assumes the values: 
$\left[R_{\Delta}\right]_{\rm H_{2}S}=3.67$ and $\left[R_{\Delta}\right]_{\rm H_{4}S_{3}}=3.53$. For the ${\rm H_{2}S}$ and ${\rm H_{4}S_{3}}$ compounds, the vertex corrections do not change $R_{\Delta}$. They do not influence the function $\Delta\left(T\right)$, which is why they are not taken into account in the further part of paper.

\subsection{The free energy, thermodynamic critical field, entropy, and the specific heat jump}

\begin{figure}
\includegraphics[width=\columnwidth]{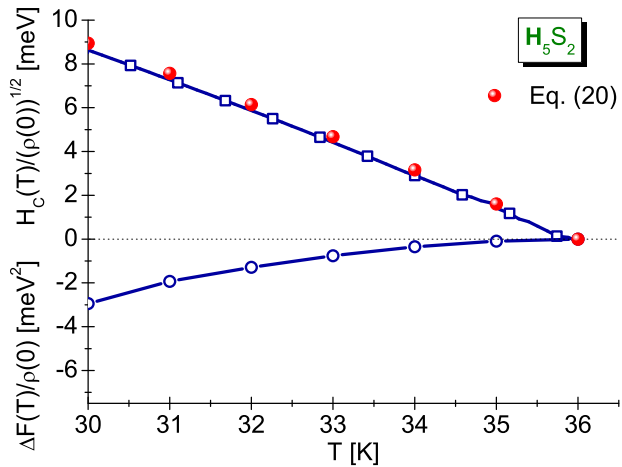}
\caption{(lower panel) The free energy difference between the superconducting and normal state as a function of temperature, and 
         (upper panel) the thermodynamic critical field. The charts were obtained as the part of classical Eliashberg formalism 
         without the vertex corrections (for the temperature area in which the results 
          obtained with the help of VECC and CEE models are identical ($T>30$~K)).}
\label{f05}
\end{figure}

The thermodynamics of superconducting state is fully determined by the dependence of order parameter on the temperature. 
Based on the results posted in \fig{f02}, it can be seen that in the temperature area $T_{C}-T \ll T_{C}$, the values of $\Delta\left(T\right)$ 
obtained for ${\rm H_{5}S_{2}}$ in the framework of VCEE and CEE schemes are physically indistinguishable. This result means that the thermodynamics of superconducting state near the critical temperature can be analyzed with the help of classical Eliashberg approach without vertex corrections.
  
The free energy difference between the superconducting and normal state has been calculated in agreement with the formula \cite{Bardeen1964A}:
\begin{eqnarray}
\label{r1-IIIC}
\frac{\Delta F}{\rho(0)} &=&-2\pi k_{B}T\sum^{M}_{m=1}
[\sqrt{\omega^2_m+\left(\Delta_m\right)^2}-|\omega_m|]\\ \nonumber
&\times&[Z^{S}_m-Z^{N}_m \frac{|\omega_m|}{\sqrt{\omega^2_m+\left(\Delta_m\right)^2}}],
\end{eqnarray}
whereas $Z^{N}_m$ and $Z^{S}_m$ denote respectively the wave function renormalization factor for the normal state ($N$) and for the superconducting state ($S$).

In Figure \ref{f05} (lower panel), we have plotted the form of function $\Delta F\left(T\right)/\rho(0)$ for ${\rm H_{5}S_{2}}$. It can be seen that the free energy difference takes negative values in the whole temperature range up to $\rm{T_{C}}$. This demonstrates thermodynamic stability of superconducting phase in the compound under investigation. For the lowest temperature taken into account, it was obtained:
$\left[\Delta F/\rho\left(0\right)\right]_{T=30\hspace{0.5mm}{\rm K}}=-2.94$~${\rm meV^2}$.
\begin{figure}
\includegraphics[width=\columnwidth]{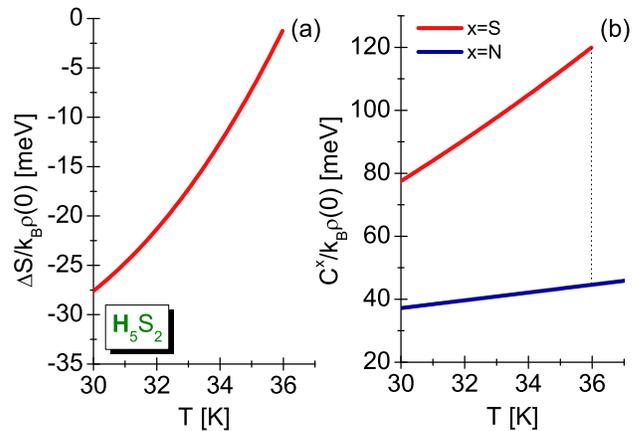}
\caption{(a) The dependence of difference in the entropy on the temperature, and (b) the specific heat in the superconducting and normal state 
             (the Eliashberg formalism without the vertex corrections).}
\label{f06}
\end{figure}

On the basis of the temperature dependence of free energy difference, it is relatively easy to determine the values of thermodynamic critical field 
($H_{C}$), the difference in entropy ($\Delta S$) between the superconducting state and the normal state, and the specific heat difference ($\Delta C$).

The thermodynamic critical field has been calculated on the basis of formula:
\begin{eqnarray}
\label{r2-IIIC}
\frac{H_{C}}{\sqrt{\rho\left(0\right)}}=\sqrt{-8\pi\frac{\Delta F}{\rho\left(0\right)}}.
\end{eqnarray}
We presented the results obtained for the ${\rm H_{5}S_{2}}$ compound in \fig{f05} (upper panel). We see that as the temperature rises, the critical field decreases so that in $T=T_C$ its value equaled zero. Assuming for the $\rm{H_{5}S_{2}}$ compound the following designation 
$H\left(0\right)=H\left(T=30\hspace{1mm}{\rm K}\right)=8.95$~meV, it is easy to see that the critical field decreases in the proportion to the square of temperature, which is well illustrated by the parabola plotted on the basis of equation \cite{Rose-Innes1969A}:
\begin{eqnarray}
\label{r3-IIIC} 
H_{C}\left(T\right)=H\left(0\right)\left[1-\left(T/T_{C}\right)^2\right].
\end{eqnarray}
The dependence $H_{C}\left(T\right)$ determined using this formula is represented by the red spheres in \fig{f05}.

The difference in the entropy between the superconducting and normal state has been estimated based on the expression:
\begin{eqnarray}
\label{r4-IIIC}
\frac{\Delta S}{k_B\rho\left(0\right)}=-\frac{d\left[\Delta F/\rho\left(0\right)\right]}{d\left(k_{B}T\right)}.
\end{eqnarray}
We presented the obtained results in \fig{f06}~(a). Physically increasing are the value of entropy up to $T_{C}$ proves the higher ordering 
of superconducting state in relation to the normal state.

The values of $\Delta C$ have been calculated using the formula:
\begin{eqnarray}
\label{r5-IIIC}
\frac{\Delta C}{k_B \rho \left(0\right)}=-\frac{1}{\beta}\frac{d^2 \left[\Delta F / \rho \left(0\right) \right]}{d \left(k_B T \right)^2}.
\end{eqnarray}
In addition, the specific heat of normal state is determined based on:
\begin{equation}
\label{r6-IIIC}
\frac{C^N}{k_B \rho(0)}=\gamma k_{B}T,
\end{equation}
where the Sommerfeld constant equals: $\gamma=\frac{2}{3}\pi^2(1+\lambda)$.
The influence of temperature on the specific heat in the superconducting state and the normal state has been presented in \fig{f06}~(b). 
The characteristic specific heat jump visible in the critical temperature is noteworthy. For the ${\rm H_{5}S_{2}}$ compound its value equals $75.34$~meV.

\begin{figure}
\includegraphics[width=\columnwidth]{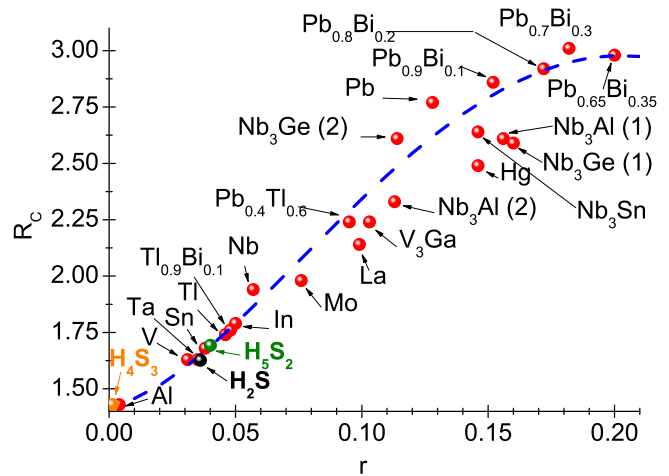}
\caption{                 The value of ratio $R_{C}$ in the dependence on the parameter $r$. 
                          The blue line represents the general trend obtained using the formula:
                          $R_{C}=1.43 \left[1+53 (r)^2 \ln(1/3r)\right]$ (CEE scheme \cite{Carbotte1990A}). } 
\label{f07}
\end{figure}

The thermodynamic parameters determined allow to calculate the dimensionless ratio:
$R_{C}=\Delta C\left(T_{C}\right)/C^{N}\left(T_{C}\right)$. As part of classic BCS theory, the quantity $R_{C}$ takes the universal value equal to $1.43$ \cite{Bardeen1957A, Bardeen1957B}. In the case of $\rm{H_5S_2}$ compound, it was obtained $R_{C}=1.69$. 
Hence, the value of $R_{C}$ for $\rm{H_{5}S_{2}}$ deviates from the predictions of BCS theory. 
Additionally, \fig{f07} presents the dependence of dimensionless parameter $R_{C}$ on $r$. The chart shows that the value of $R_{C}$ obtained for  
$\rm{H_{5}S_{2}}$ perfectly fits the general trend anticipated by the classic Eliashberg formalism. 

In the case of ${\rm H_{2}S}$ compound the ratio $R_{C}=1.63$ is close to $\left[R_{C}\right]_{\rm{H_{5}S_{2}}}$. 
The parameter $R_{C}$ for $\rm{H_{4}S_{3}}$ is equal to the value predicted by the BCS theory.

\section{SUMMARY}

We have calculated the thermodynamic parameters of superconducting state for the ${\rm H_{5}S_{2}}$ compound under the pressure at $112$~GPa. 
We have solved the Eliashberg equations on the imaginary axis, both including the first-order vertex corrections to the electron-phonon interaction, 
as well as without vertex corrections. For both cases, we have obtained anomalously high values of Coulomb pseudopotential: $\left[\mu^{\star}\left(\omega^{\left(1\right)}_{c}\right)\right]_{VCEE}=0.589$ and 
$\left[\mu^{\star}\left(\omega^{\left(1\right)}_{c}\right)\right]_{CEE}=0.402$, while $\omega^{\left(1\right)}_{c}=3\omega_{D}$. 
Note that increase in the cut-off frequency only increases value of $\mu^{\star}$ ({\it e.g.} for VCEE case $\mu^{\star}=1.241$, where 
$\omega^{\left(2\right)}_{c}=10\omega_{D}$). The calculated values of $\mu^{\star}$ proves that this parameter cannot be associated only with the depairing Coulomb correlations - in principle, it should be treated as the effective parameter of fitting the model to experimental data. 
The above results mean that it is very unlikely that the low-temperature superconducting state of compressed sulfur-hydrogen system is induced in the ${\rm H_{5}S_{2}}$ compound. In our opinion, experimentally was observed the superconducting state in the ${\rm H_{2}S}$ compound, which is kinetically protected in the samples prepared at the low temperature \cite{Li2014A, Li2016A}. It should be emphasized that in the case of ${\rm H_{2}S}$ reproducing the experimental dependence of critical temperature on the pressure does not require anomalously high value of Coulomb pseudopotential  \cite{Durajski2015A}. 

In our paper, we also analyzed the thermodynamic properties of superconducting state in the ${\rm H_{4}S_{3}}$ compound. 
It is characterized by the very low value of critical temperature ($\left[T_{C}\right]_{\rm max}\sim 2$~K) \cite{Li2016A}, and it cannot be 
associated with the low temperature superconducting state of compressed dihydrogen sulfide. Probably, we have the analogous situation here like 
for the ${\rm H_{5}S_{2}}$ compound, where kinetically protected ${\rm H_{2}S}$ in the samples prepared at the low temperature is responsible for the observed superconductivity.

As part of the analysis, we calculated the thermodynamic parameters of superconducting state for the ${\rm H_{5}S_{2}}$, 
${\rm H_{2}S}$, and ${\rm H_{4}S_{3}}$ compounds. In the case of ${\rm H_{5}S_{2}}$, we have shown that the first order vertex corrections to the electron-phonon interaction play the important role, {\it i.e.} they significantly change the thermodynamics of superconducting state in the low temperature range, while nearby $T_{C}$ they are practically irrelevant. For this reason, the dimensionless parameter $R_{\Delta}$ in the VCEE scheme is equal to $4.85$, and in the CEE scheme it equals $3.77$. On the other hand, for both cases it was obtained $R_{C}=1.693$. The above values prove that the superconducting state in the ${\rm{H_{5}S_{2}}}$ compound is not the state of BCS type. The superconducting state for  
${\rm H_{2}S}$ has the thermodynamic parameters with values are close to the ones determined for ${\rm H_{5}S_{2}}$ in the CEE scheme. 
In particular, we got: $R_{\Delta}=3.67$ and $R_{C}=1.626$. On the other hand, the superconducting state of ${\rm H_{4}S_{3}}$ compound is the BCS type.

Regarding the results presented by Ishikawa {\it et al.} \cite{Ishikawa2016A}, in respect to the Eliashberg function (harmonic limit), 
it should be assumed that these are the results proving the impossibility of correct description of the low-temperature superconducting phase 
of compressed sulfur-hydrogen system.

\section{Author Contributions}

Ma{\l}gorzata Kostrzewa wrote the code for numerical calculations (VCEE scheme), carried out the calculations (VCEE scheme), 
                        and participated in writing the manuscript.
R. Szcz{\c{e}}{\'s}niak wrote the code for numerical calculations (CEE scheme) and participated in writing the manuscript.
Joanna K. Kalaga collected data and drafted the final version of the manuscript.
Izabela A. Wrona carried out the calculations (CEE scheme).
All authors reviewed the manuscript.

\section{Additional Information}

{\bf Competing Interests:} The authors declare no competing interests.



\begin{thebibliography}{65}
\expandafter\ifx\csname natexlab\endcsname\relax\def\natexlab#1{#1}\fi
\expandafter\ifx\csname bibnamefont\endcsname\relax
  \def\bibnamefont#1{#1}\fi
\expandafter\ifx\csname bibfnamefont\endcsname\relax
  \def\bibfnamefont#1{#1}\fi
\expandafter\ifx\csname citenamefont\endcsname\relax
  \def\citenamefont#1{#1}\fi
\expandafter\ifx\csname url\endcsname\relax
  \def\url#1{\texttt{#1}}\fi
\expandafter\ifx\csname urlprefix\endcsname\relax\def\urlprefix{URL }\fi
\providecommand{\bibinfo}[2]{#2}
\providecommand{\eprint}[2][]{\url{#2}}

\bibitem[{\citenamefont{Eliashberg}(1960)}]{Eliashberg1960A}
\bibinfo{author}{\bibfnamefont{G.~M.} \bibnamefont{Eliashberg}},
  \bibinfo{journal}{Soviet Physics JETP} \textbf{\bibinfo{volume}{11}},
  \bibinfo{pages}{696} (\bibinfo{year}{1960}).

\bibitem[{\citenamefont{Carbotte}(1990)}]{Carbotte1990A}
\bibinfo{author}{\bibfnamefont{J.~P.} \bibnamefont{Carbotte}},
  \bibinfo{journal}{Reviews of Modern Physics} \textbf{\bibinfo{volume}{62}},
  \bibinfo{pages}{1027} (\bibinfo{year}{1990}).

\bibitem[{\citenamefont{Carbotte and Marsiglio}(2003)}]{Carbotte2003A}
\bibinfo{author}{\bibfnamefont{J.~P.} \bibnamefont{Carbotte}} \bibnamefont{and}
  \bibinfo{author}{\bibfnamefont{F.}~\bibnamefont{Marsiglio}}, in
  \emph{\bibinfo{booktitle}{The Physics of Superconductors edited by K. H.
  Bennemann and J. B. Ketterson}} (\bibinfo{publisher}{Springer Berlin
  Heidelberg}, \bibinfo{year}{2003}).

\bibitem[{\citenamefont{Morel and Anderson}(1962)}]{Morel1962A}
\bibinfo{author}{\bibfnamefont{P.}~\bibnamefont{Morel}} \bibnamefont{and}
  \bibinfo{author}{\bibfnamefont{P.~W.} \bibnamefont{Anderson}},
  \bibinfo{journal}{Physical Review} \textbf{\bibinfo{volume}{125}},
  \bibinfo{pages}{1263} (\bibinfo{year}{1962}).

\bibitem[{\citenamefont{Bauer et~al.}(2012)\citenamefont{Bauer, Han, and
  Gunnarsson}}]{Bauer2012A}
\bibinfo{author}{\bibfnamefont{J.}~\bibnamefont{Bauer}},
  \bibinfo{author}{\bibfnamefont{J.~E.} \bibnamefont{Han}}, \bibnamefont{and}
  \bibinfo{author}{\bibfnamefont{O.}~\bibnamefont{Gunnarsson}},
  \bibinfo{journal}{Journal of Physics: Condensed Matter}
  \textbf{\bibinfo{volume}{24}}, \bibinfo{pages}{492202}
  (\bibinfo{year}{2012}).

\bibitem[{\citenamefont{Baroni et~al.}(1986)\citenamefont{Baroni, Corso,
  de~Gironcoli, Giannozzi, Cavazzoni, Ballabio, Scandolo, Chiarotti, Focher,
  Pasquarello et~al.}}]{Baroni1986A}
\bibinfo{author}{\bibfnamefont{S.}~\bibnamefont{Baroni}},
  \bibinfo{author}{\bibfnamefont{A.~D.} \bibnamefont{Corso}},
  \bibinfo{author}{\bibfnamefont{S.}~\bibnamefont{de~Gironcoli}},
  \bibinfo{author}{\bibfnamefont{P.}~\bibnamefont{Giannozzi}},
  \bibinfo{author}{\bibfnamefont{C.}~\bibnamefont{Cavazzoni}},
  \bibinfo{author}{\bibfnamefont{G.}~\bibnamefont{Ballabio}},
  \bibinfo{author}{\bibfnamefont{S.}~\bibnamefont{Scandolo}},
  \bibinfo{author}{\bibfnamefont{G.}~\bibnamefont{Chiarotti}},
  \bibinfo{author}{\bibfnamefont{P.}~\bibnamefont{Focher}},
  \bibinfo{author}{\bibfnamefont{A.}~\bibnamefont{Pasquarello}},
  \bibnamefont{et~al.}, \emph{\bibinfo{title}{Quantum espresso}}
  (\bibinfo{year}{1986}), \urlprefix\url{http://www.pwscf.org.}

\bibitem[{\citenamefont{Giannozzi et~al.}(2009)\citenamefont{Giannozzi, Baroni,
  Bonini, Calandra, Car, Cavazzoni, Ceresoli, Chiarotti, Cococcioni, Dabo
  et~al.}}]{Giannozzi2009A}
\bibinfo{author}{\bibfnamefont{P.}~\bibnamefont{Giannozzi}},
  \bibinfo{author}{\bibfnamefont{S.}~\bibnamefont{Baroni}},
  \bibinfo{author}{\bibfnamefont{N.}~\bibnamefont{Bonini}},
  \bibinfo{author}{\bibfnamefont{M.}~\bibnamefont{Calandra}},
  \bibinfo{author}{\bibfnamefont{R.}~\bibnamefont{Car}},
  \bibinfo{author}{\bibfnamefont{C.}~\bibnamefont{Cavazzoni}},
  \bibinfo{author}{\bibfnamefont{D.}~\bibnamefont{Ceresoli}},
  \bibinfo{author}{\bibfnamefont{G.~L.} \bibnamefont{Chiarotti}},
  \bibinfo{author}{\bibfnamefont{M.}~\bibnamefont{Cococcioni}},
  \bibinfo{author}{\bibfnamefont{I.}~\bibnamefont{Dabo}}, \bibnamefont{et~al.},
  \bibinfo{journal}{Journal of Physics: Condensed Matter}
  \textbf{\bibinfo{volume}{21}}, \bibinfo{pages}{395502}
  (\bibinfo{year}{2009}).

\bibitem[{\citenamefont{Kim and Tesanovic}(1993)}]{Kim1993A}
\bibinfo{author}{\bibfnamefont{J.~H.} \bibnamefont{Kim}} \bibnamefont{and}
  \bibinfo{author}{\bibfnamefont{Z.}~\bibnamefont{Tesanovic}},
  \bibinfo{journal}{Physical Review Letters} \textbf{\bibinfo{volume}{71}},
  \bibinfo{pages}{4218} (\bibinfo{year}{1993}).

\bibitem[{\citenamefont{Drozdov et~al.}(2014)\citenamefont{Drozdov, Eremets,
  and Troyan}}]{Drozdov2014A}
\bibinfo{author}{\bibfnamefont{A.~P.} \bibnamefont{Drozdov}},
  \bibinfo{author}{\bibfnamefont{M.~I.} \bibnamefont{Eremets}},
  \bibnamefont{and} \bibinfo{author}{\bibfnamefont{I.~A.}
  \bibnamefont{Troyan}}, \bibinfo{journal}{arXiv: 1412.0460}
  (\bibinfo{year}{2014}).

\bibitem[{\citenamefont{Drozdov et~al.}(2015)\citenamefont{Drozdov, Eremets,
  Troyan, Ksenofontov, and Shylin}}]{Drozdov2015A}
\bibinfo{author}{\bibfnamefont{A.~P.} \bibnamefont{Drozdov}},
  \bibinfo{author}{\bibfnamefont{M.~I.} \bibnamefont{Eremets}},
  \bibinfo{author}{\bibfnamefont{I.~A.} \bibnamefont{Troyan}},
  \bibinfo{author}{\bibfnamefont{V.}~\bibnamefont{Ksenofontov}},
  \bibnamefont{and} \bibinfo{author}{\bibfnamefont{S.~I.}
  \bibnamefont{Shylin}}, \bibinfo{journal}{Nature}
  \textbf{\bibinfo{volume}{525}}, \bibinfo{pages}{73} (\bibinfo{year}{2015}).

\bibitem[{\citenamefont{Li et~al.}(2014)\citenamefont{Li, Hao, Liu, Li, and
  Ma}}]{Li2014A}
\bibinfo{author}{\bibfnamefont{Y.}~\bibnamefont{Li}},
  \bibinfo{author}{\bibfnamefont{J.}~\bibnamefont{Hao}},
  \bibinfo{author}{\bibfnamefont{H.}~\bibnamefont{Liu}},
  \bibinfo{author}{\bibfnamefont{Y.}~\bibnamefont{Li}}, \bibnamefont{and}
  \bibinfo{author}{\bibfnamefont{Y.}~\bibnamefont{Ma}}, \bibinfo{journal}{The
  Journal of Chemical Physics} \textbf{\bibinfo{volume}{140}},
  \bibinfo{pages}{174712} (\bibinfo{year}{2014}).

\bibitem[{\citenamefont{Endo et~al.}(1996)\citenamefont{Endo, Honda, Sasaki,
  Shimizu, Shimomura, and Kikegawa}}]{Endo1996A}
\bibinfo{author}{\bibfnamefont{S.}~\bibnamefont{Endo}},
  \bibinfo{author}{\bibfnamefont{A.}~\bibnamefont{Honda}},
  \bibinfo{author}{\bibfnamefont{S.}~\bibnamefont{Sasaki}},
  \bibinfo{author}{\bibfnamefont{H.}~\bibnamefont{Shimizu}},
  \bibinfo{author}{\bibfnamefont{O.}~\bibnamefont{Shimomura}},
  \bibnamefont{and} \bibinfo{author}{\bibfnamefont{T.}~\bibnamefont{Kikegawa}},
  \bibinfo{journal}{Physical Review B} \textbf{\bibinfo{volume}{54}},
  \bibinfo{pages}{R717(R)} (\bibinfo{year}{1996}).

\bibitem[{\citenamefont{Fujihisa et~al.}(1998)\citenamefont{Fujihisa, Yamawaki,
  Sakashita, Aoki, Sasaki, and Shimizu}}]{Fujihisa1998A}
\bibinfo{author}{\bibfnamefont{H.}~\bibnamefont{Fujihisa}},
  \bibinfo{author}{\bibfnamefont{H.}~\bibnamefont{Yamawaki}},
  \bibinfo{author}{\bibfnamefont{M.}~\bibnamefont{Sakashita}},
  \bibinfo{author}{\bibfnamefont{K.}~\bibnamefont{Aoki}},
  \bibinfo{author}{\bibfnamefont{S.}~\bibnamefont{Sasaki}}, \bibnamefont{and}
  \bibinfo{author}{\bibfnamefont{H.}~\bibnamefont{Shimizu}},
  \bibinfo{journal}{Physical Review B} \textbf{\bibinfo{volume}{57}},
  \bibinfo{pages}{2651} (\bibinfo{year}{1998}).

\bibitem[{\citenamefont{Rousseau et~al.}(2000)\citenamefont{Rousseau, Boero,
  Bernasconi, Parrinello, and Terakura}}]{Rousseau2000A}
\bibinfo{author}{\bibfnamefont{R.}~\bibnamefont{Rousseau}},
  \bibinfo{author}{\bibfnamefont{M.}~\bibnamefont{Boero}},
  \bibinfo{author}{\bibfnamefont{M.}~\bibnamefont{Bernasconi}},
  \bibinfo{author}{\bibfnamefont{M.}~\bibnamefont{Parrinello}},
  \bibnamefont{and} \bibinfo{author}{\bibfnamefont{K.}~\bibnamefont{Terakura}},
  \bibinfo{journal}{Physical Review Letters} \textbf{\bibinfo{volume}{85}},
  \bibinfo{pages}{1254} (\bibinfo{year}{2000}).

\bibitem[{\citenamefont{Sakashita et~al.}(2000)\citenamefont{Sakashita,
  Fujihisa, Yamawaki, and Aoki}}]{Sakashita2000A}
\bibinfo{author}{\bibfnamefont{M.}~\bibnamefont{Sakashita}},
  \bibinfo{author}{\bibfnamefont{H.}~\bibnamefont{Fujihisa}},
  \bibinfo{author}{\bibfnamefont{H.}~\bibnamefont{Yamawaki}}, \bibnamefont{and}
  \bibinfo{author}{\bibfnamefont{K.}~\bibnamefont{Aoki}}, \bibinfo{journal}{J.
  Phys. Chem. A} \textbf{\bibinfo{volume}{104}}, \bibinfo{pages}{8838}
  (\bibinfo{year}{2000}).

\bibitem[{\citenamefont{Fujihisa et~al.}(2004)\citenamefont{Fujihisa, Yamawaki,
  Sakashita, Nakayama, Yamada, and Aoki}}]{Fujihisa2004A}
\bibinfo{author}{\bibfnamefont{H.}~\bibnamefont{Fujihisa}},
  \bibinfo{author}{\bibfnamefont{H.}~\bibnamefont{Yamawaki}},
  \bibinfo{author}{\bibfnamefont{M.}~\bibnamefont{Sakashita}},
  \bibinfo{author}{\bibfnamefont{A.}~\bibnamefont{Nakayama}},
  \bibinfo{author}{\bibfnamefont{T.}~\bibnamefont{Yamada}}, \bibnamefont{and}
  \bibinfo{author}{\bibfnamefont{K.}~\bibnamefont{Aoki}},
  \bibinfo{journal}{Physical Review B} \textbf{\bibinfo{volume}{69}},
  \bibinfo{pages}{214102} (\bibinfo{year}{2004}).

\bibitem[{\citenamefont{Sakashita et~al.}(1997)\citenamefont{Sakashita,
  Yamawaki, Fujihisa, Aoki, Sasaki, and Shimizu}}]{Sakashita1997A}
\bibinfo{author}{\bibfnamefont{M.}~\bibnamefont{Sakashita}},
  \bibinfo{author}{\bibfnamefont{H.}~\bibnamefont{Yamawaki}},
  \bibinfo{author}{\bibfnamefont{H.}~\bibnamefont{Fujihisa}},
  \bibinfo{author}{\bibfnamefont{K.}~\bibnamefont{Aoki}},
  \bibinfo{author}{\bibfnamefont{S.}~\bibnamefont{Sasaki}}, \bibnamefont{and}
  \bibinfo{author}{\bibfnamefont{H.}~\bibnamefont{Shimizu}},
  \bibinfo{journal}{Physical Review Letters} \textbf{\bibinfo{volume}{79}},
  \bibinfo{pages}{1082} (\bibinfo{year}{1997}).

\bibitem[{\citenamefont{Shimizu et~al.}(1991)\citenamefont{Shimizu, Nakamichi,
  and Sasaki}}]{Shimizu1991A}
\bibinfo{author}{\bibfnamefont{H.}~\bibnamefont{Shimizu}},
  \bibinfo{author}{\bibfnamefont{Y.}~\bibnamefont{Nakamichi}},
  \bibnamefont{and} \bibinfo{author}{\bibfnamefont{S.}~\bibnamefont{Sasaki}},
  \bibinfo{journal}{The Journal of Chemical Physics}
  \textbf{\bibinfo{volume}{95}}, \bibinfo{pages}{2036} (\bibinfo{year}{1991}).

\bibitem[{\citenamefont{Endo et~al.}(1994)\citenamefont{Endo, Ichimiya, Koto,
  Sasaki, and Shimizui}}]{Endo1994A}
\bibinfo{author}{\bibfnamefont{S.}~\bibnamefont{Endo}},
  \bibinfo{author}{\bibfnamefont{N.}~\bibnamefont{Ichimiya}},
  \bibinfo{author}{\bibfnamefont{K.}~\bibnamefont{Koto}},
  \bibinfo{author}{\bibfnamefont{S.}~\bibnamefont{Sasaki}}, \bibnamefont{and}
  \bibinfo{author}{\bibfnamefont{H.}~\bibnamefont{Shimizui}},
  \bibinfo{journal}{Physical Review B} \textbf{\bibinfo{volume}{50}},
  \bibinfo{pages}{5865} (\bibinfo{year}{1994}).

\bibitem[{\citenamefont{Endo et~al.}(1998)\citenamefont{Endo, Honda, Koto,
  Shimomura, Kikegawa, and Hamaya}}]{Endo1998A}
\bibinfo{author}{\bibfnamefont{S.}~\bibnamefont{Endo}},
  \bibinfo{author}{\bibfnamefont{A.}~\bibnamefont{Honda}},
  \bibinfo{author}{\bibfnamefont{K.}~\bibnamefont{Koto}},
  \bibinfo{author}{\bibfnamefont{O.}~\bibnamefont{Shimomura}},
  \bibinfo{author}{\bibfnamefont{T.}~\bibnamefont{Kikegawa}}, \bibnamefont{and}
  \bibinfo{author}{\bibfnamefont{N.}~\bibnamefont{Hamaya}},
  \bibinfo{journal}{Physical Review B} \textbf{\bibinfo{volume}{57}},
  \bibinfo{pages}{5699} (\bibinfo{year}{1998}).

\bibitem[{\citenamefont{Shimizu and Sasaki}(1992)}]{Shimizu1992A}
\bibinfo{author}{\bibfnamefont{H.}~\bibnamefont{Shimizu}} \bibnamefont{and}
  \bibinfo{author}{\bibfnamefont{S.}~\bibnamefont{Sasaki}},
  \bibinfo{journal}{Science} \textbf{\bibinfo{volume}{257}},
  \bibinfo{pages}{514} (\bibinfo{year}{1992}).

\bibitem[{\citenamefont{Shimizu et~al.}(1995)\citenamefont{Shimizu, Yamaguchi,
  Sasaki, Honda, Endo, and Kobayashi}}]{Shimizu1995A}
\bibinfo{author}{\bibfnamefont{H.}~\bibnamefont{Shimizu}},
  \bibinfo{author}{\bibfnamefont{H.}~\bibnamefont{Yamaguchi}},
  \bibinfo{author}{\bibfnamefont{S.}~\bibnamefont{Sasaki}},
  \bibinfo{author}{\bibfnamefont{A.}~\bibnamefont{Honda}},
  \bibinfo{author}{\bibfnamefont{S.}~\bibnamefont{Endo}}, \bibnamefont{and}
  \bibinfo{author}{\bibfnamefont{M.}~\bibnamefont{Kobayashi}},
  \bibinfo{journal}{Physical Review B} \textbf{\bibinfo{volume}{51}},
  \bibinfo{pages}{9391} (\bibinfo{year}{1995}).

\bibitem[{\citenamefont{Shimizu et~al.}(1997)\citenamefont{Shimizu, Ushida,
  Sasaki, Sakashita, Yamawaki, and Aoki}}]{Shimizu1997A}
\bibinfo{author}{\bibfnamefont{H.}~\bibnamefont{Shimizu}},
  \bibinfo{author}{\bibfnamefont{T.}~\bibnamefont{Ushida}},
  \bibinfo{author}{\bibfnamefont{S.}~\bibnamefont{Sasaki}},
  \bibinfo{author}{\bibfnamefont{M.}~\bibnamefont{Sakashita}},
  \bibinfo{author}{\bibfnamefont{H.}~\bibnamefont{Yamawaki}}, \bibnamefont{and}
  \bibinfo{author}{\bibfnamefont{K.}~\bibnamefont{Aoki}},
  \bibinfo{journal}{Physical Review B} \textbf{\bibinfo{volume}{55}},
  \bibinfo{pages}{5538} (\bibinfo{year}{1997}).

\bibitem[{\citenamefont{Loveday et~al.}(2000)\citenamefont{Loveday, Nelmes,
  Klotz, Besson, and Hamel}}]{Loveday2000A}
\bibinfo{author}{\bibfnamefont{J.~S.} \bibnamefont{Loveday}},
  \bibinfo{author}{\bibfnamefont{R.~J.} \bibnamefont{Nelmes}},
  \bibinfo{author}{\bibfnamefont{S.}~\bibnamefont{Klotz}},
  \bibinfo{author}{\bibfnamefont{J.~M.} \bibnamefont{Besson}},
  \bibnamefont{and} \bibinfo{author}{\bibfnamefont{G.}~\bibnamefont{Hamel}},
  \bibinfo{journal}{Physical Review Letters} \textbf{\bibinfo{volume}{85}},
  \bibinfo{pages}{1024} (\bibinfo{year}{2000}).

\bibitem[{\citenamefont{Kume et~al.}(2002)\citenamefont{Kume, Fukaya, Sasaki,
  and Shimizu}}]{Kume2002A}
\bibinfo{author}{\bibfnamefont{T.}~\bibnamefont{Kume}},
  \bibinfo{author}{\bibfnamefont{Y.}~\bibnamefont{Fukaya}},
  \bibinfo{author}{\bibfnamefont{S.}~\bibnamefont{Sasaki}}, \bibnamefont{and}
  \bibinfo{author}{\bibfnamefont{H.}~\bibnamefont{Shimizu}},
  \bibinfo{journal}{Review of Scientific Instruments}
  \textbf{\bibinfo{volume}{73}}, \bibinfo{pages}{2355} (\bibinfo{year}{2002}).

\bibitem[{\citenamefont{Durajski et~al.}(2015)\citenamefont{Durajski,
  Szcz{\c{e}}{\'s}niak, and Li}}]{Durajski2015A}
\bibinfo{author}{\bibfnamefont{A.~P.} \bibnamefont{Durajski}},
  \bibinfo{author}{\bibfnamefont{R.}~\bibnamefont{Szcz{\c{e}}{\'s}niak}},
  \bibnamefont{and} \bibinfo{author}{\bibfnamefont{Y.}~\bibnamefont{Li}},
  \bibinfo{journal}{Physica C} \textbf{\bibinfo{volume}{515}},
  \bibinfo{pages}{1} (\bibinfo{year}{2015}).

\bibitem[{\citenamefont{Ishikawa et~al.}(2016)\citenamefont{Ishikawa,
  Nakanishi, Shimizu, Katayama-Yoshida, Oda, and Suzuki}}]{Ishikawa2016A}
\bibinfo{author}{\bibfnamefont{T.}~\bibnamefont{Ishikawa}},
  \bibinfo{author}{\bibfnamefont{A.}~\bibnamefont{Nakanishi}},
  \bibinfo{author}{\bibfnamefont{K.}~\bibnamefont{Shimizu}},
  \bibinfo{author}{\bibfnamefont{H.}~\bibnamefont{Katayama-Yoshida}},
  \bibinfo{author}{\bibfnamefont{T.}~\bibnamefont{Oda}}, \bibnamefont{and}
  \bibinfo{author}{\bibfnamefont{N.}~\bibnamefont{Suzuki}},
  \bibinfo{journal}{Scientific Reports} \textbf{\bibinfo{volume}{6}},
  \bibinfo{pages}{23160} (\bibinfo{year}{2016}).

\bibitem[{\citenamefont{Errea et~al.}(2015)\citenamefont{Errea, Calandra,
  Pickard, Richard, Needs, Li, Liu, Zhang, Ma, and Mauri}}]{Errea2015A}
\bibinfo{author}{\bibfnamefont{I.}~\bibnamefont{Errea}},
  \bibinfo{author}{\bibfnamefont{M.}~\bibnamefont{Calandra}},
  \bibinfo{author}{\bibfnamefont{C.~J.} \bibnamefont{Pickard}},
  \bibinfo{author}{\bibfnamefont{J.~N.} \bibnamefont{Richard}},
  \bibinfo{author}{\bibfnamefont{J.}~\bibnamefont{Needs}},
  \bibinfo{author}{\bibfnamefont{Y.}~\bibnamefont{Li}},
  \bibinfo{author}{\bibfnamefont{H.}~\bibnamefont{Liu}},
  \bibinfo{author}{\bibfnamefont{Y.}~\bibnamefont{Zhang}},
  \bibinfo{author}{\bibfnamefont{Y.}~\bibnamefont{Ma}}, \bibnamefont{and}
  \bibinfo{author}{\bibfnamefont{F.}~\bibnamefont{Mauri}},
  \bibinfo{journal}{Physical Review Letters} \textbf{\bibinfo{volume}{114}},
  \bibinfo{pages}{157004} (\bibinfo{year}{2015}).

\bibitem[{\citenamefont{Errea et~al.}(2016)\citenamefont{Errea, Calandra,
  Pickard, Nelson, Needs, Li, Liu, Zhang, Ma, and Mauri}}]{Errea2016A}
\bibinfo{author}{\bibfnamefont{I.}~\bibnamefont{Errea}},
  \bibinfo{author}{\bibfnamefont{M.}~\bibnamefont{Calandra}},
  \bibinfo{author}{\bibfnamefont{C.~J.} \bibnamefont{Pickard}},
  \bibinfo{author}{\bibfnamefont{J.~R.} \bibnamefont{Nelson}},
  \bibinfo{author}{\bibfnamefont{R.~J.} \bibnamefont{Needs}},
  \bibinfo{author}{\bibfnamefont{Y.}~\bibnamefont{Li}},
  \bibinfo{author}{\bibfnamefont{H.}~\bibnamefont{Liu}},
  \bibinfo{author}{\bibfnamefont{Y.}~\bibnamefont{Zhang}},
  \bibinfo{author}{\bibfnamefont{Y.}~\bibnamefont{Ma}}, \bibnamefont{and}
  \bibinfo{author}{\bibfnamefont{F.}~\bibnamefont{Mauri}},
  \bibinfo{journal}{Nature} \textbf{\bibinfo{volume}{532}}, \bibinfo{pages}{81}
  (\bibinfo{year}{2016}).

\bibitem[{\citenamefont{Li et~al.}(2016)\citenamefont{Li, Wang, Liu, Zhang,
  Hao, Pickard, Nelson, Needs, Li, Huang et~al.}}]{Li2016A}
\bibinfo{author}{\bibfnamefont{Y.}~\bibnamefont{Li}},
  \bibinfo{author}{\bibfnamefont{L.}~\bibnamefont{Wang}},
  \bibinfo{author}{\bibfnamefont{H.}~\bibnamefont{Liu}},
  \bibinfo{author}{\bibfnamefont{Y.}~\bibnamefont{Zhang}},
  \bibinfo{author}{\bibfnamefont{J.}~\bibnamefont{Hao}},
  \bibinfo{author}{\bibfnamefont{C.~J.} \bibnamefont{Pickard}},
  \bibinfo{author}{\bibfnamefont{J.~R.} \bibnamefont{Nelson}},
  \bibinfo{author}{\bibfnamefont{R.~J.} \bibnamefont{Needs}},
  \bibinfo{author}{\bibfnamefont{W.}~\bibnamefont{Li}},
  \bibinfo{author}{\bibfnamefont{Y.}~\bibnamefont{Huang}},
  \bibnamefont{et~al.}, \bibinfo{journal}{Physical Review B}
  \textbf{\bibinfo{volume}{93}}, \bibinfo{pages}{020103(R)}
  (\bibinfo{year}{2016}).

\bibitem[{\citenamefont{Duan et~al.}(2015)\citenamefont{Duan, Huang, Tian, Li,
  Yu, Liu, Ma, Liu, and Cui}}]{Duan2015A}
\bibinfo{author}{\bibfnamefont{D.}~\bibnamefont{Duan}},
  \bibinfo{author}{\bibfnamefont{X.}~\bibnamefont{Huang}},
  \bibinfo{author}{\bibfnamefont{F.}~\bibnamefont{Tian}},
  \bibinfo{author}{\bibfnamefont{D.}~\bibnamefont{Li}},
  \bibinfo{author}{\bibfnamefont{H.}~\bibnamefont{Yu}},
  \bibinfo{author}{\bibfnamefont{Y.}~\bibnamefont{Liu}},
  \bibinfo{author}{\bibfnamefont{Y.}~\bibnamefont{Ma}},
  \bibinfo{author}{\bibfnamefont{B.}~\bibnamefont{Liu}}, \bibnamefont{and}
  \bibinfo{author}{\bibfnamefont{T.}~\bibnamefont{Cui}},
  \bibinfo{journal}{Physical Review B} \textbf{\bibinfo{volume}{91}},
  \bibinfo{pages}{180502(R)} (\bibinfo{year}{2015}).

\bibitem[{\citenamefont{Bernstein et~al.}(2015)\citenamefont{Bernstein,
  Hellberg, Johannes, Mazin, and Mehl}}]{Bernstein2015A}
\bibinfo{author}{\bibfnamefont{N.}~\bibnamefont{Bernstein}},
  \bibinfo{author}{\bibfnamefont{C.~S.} \bibnamefont{Hellberg}},
  \bibinfo{author}{\bibfnamefont{M.~D.} \bibnamefont{Johannes}},
  \bibinfo{author}{\bibfnamefont{I.~I.} \bibnamefont{Mazin}}, \bibnamefont{and}
  \bibinfo{author}{\bibfnamefont{M.~J.} \bibnamefont{Mehl}},
  \bibinfo{journal}{Physical Review B} \textbf{\bibinfo{volume}{91}},
  \bibinfo{pages}{060511(R)} (\bibinfo{year}{2015}).

\bibitem[{\citenamefont{Einaga et~al.}(2016)\citenamefont{Einaga, Sakata,
  Ishikawa, Shimizu, Eremets, Drozdov, Troyan, Hirao, and
  Ohishi}}]{Einaga2016A}
\bibinfo{author}{\bibfnamefont{M.}~\bibnamefont{Einaga}},
  \bibinfo{author}{\bibfnamefont{M.}~\bibnamefont{Sakata}},
  \bibinfo{author}{\bibfnamefont{T.}~\bibnamefont{Ishikawa}},
  \bibinfo{author}{\bibfnamefont{K.}~\bibnamefont{Shimizu}},
  \bibinfo{author}{\bibfnamefont{M.~I.} \bibnamefont{Eremets}},
  \bibinfo{author}{\bibfnamefont{A.~P.} \bibnamefont{Drozdov}},
  \bibinfo{author}{\bibfnamefont{I.~A.} \bibnamefont{Troyan}},
  \bibinfo{author}{\bibfnamefont{N.}~\bibnamefont{Hirao}}, \bibnamefont{and}
  \bibinfo{author}{\bibfnamefont{Y.}~\bibnamefont{Ohishi}},
  \bibinfo{journal}{Nature Physics} \textbf{\bibinfo{volume}{12}},
  \bibinfo{pages}{835} (\bibinfo{year}{2016}).

\bibitem[{\citenamefont{Duan et~al.}(2014)\citenamefont{Duan, Liu, Tian, Li,
  Huang, Zhao, Yu, Liu, Tian, and Cui}}]{Duan2014A}
\bibinfo{author}{\bibfnamefont{D.}~\bibnamefont{Duan}},
  \bibinfo{author}{\bibfnamefont{Y.}~\bibnamefont{Liu}},
  \bibinfo{author}{\bibfnamefont{F.}~\bibnamefont{Tian}},
  \bibinfo{author}{\bibfnamefont{D.}~\bibnamefont{Li}},
  \bibinfo{author}{\bibfnamefont{X.}~\bibnamefont{Huang}},
  \bibinfo{author}{\bibfnamefont{Z.}~\bibnamefont{Zhao}},
  \bibinfo{author}{\bibfnamefont{H.}~\bibnamefont{Yu}},
  \bibinfo{author}{\bibfnamefont{B.}~\bibnamefont{Liu}},
  \bibinfo{author}{\bibfnamefont{W.}~\bibnamefont{Tian}}, \bibnamefont{and}
  \bibinfo{author}{\bibfnamefont{T.}~\bibnamefont{Cui}},
  \bibinfo{journal}{Scientific Reports} \textbf{\bibinfo{volume}{4}},
  \bibinfo{pages}{6968} (\bibinfo{year}{2014}).

\bibitem[{\citenamefont{Durajski}(2016)}]{Durajski2016A}
\bibinfo{author}{\bibfnamefont{A.~P.} \bibnamefont{Durajski}},
  \bibinfo{journal}{Scientific Reports} \textbf{\bibinfo{volume}{6}},
  \bibinfo{pages}{38570} (\bibinfo{year}{2016}).

\bibitem[{\citenamefont{Sano et~al.}(2016)\citenamefont{Sano, Koretsune,
  Tadano, Akashi, and Arita}}]{Sano2016A}
\bibinfo{author}{\bibfnamefont{W.}~\bibnamefont{Sano}},
  \bibinfo{author}{\bibfnamefont{T.}~\bibnamefont{Koretsune}},
  \bibinfo{author}{\bibfnamefont{T.}~\bibnamefont{Tadano}},
  \bibinfo{author}{\bibfnamefont{R.}~\bibnamefont{Akashi}}, \bibnamefont{and}
  \bibinfo{author}{\bibfnamefont{R.}~\bibnamefont{Arita}},
  \bibinfo{journal}{Physical Review B} \textbf{\bibinfo{volume}{93}},
  \bibinfo{pages}{094525} (\bibinfo{year}{2016}).

\bibitem[{\citenamefont{Wei et~al.}(2010)\citenamefont{Wei, Jiang-Long,
  Liang-Jian, and Zhi}}]{Wei2010A}
\bibinfo{author}{\bibfnamefont{F.}~\bibnamefont{Wei}},
  \bibinfo{author}{\bibfnamefont{W.}~\bibnamefont{Jiang-Long}},
  \bibinfo{author}{\bibfnamefont{Z.}~\bibnamefont{Liang-Jian}},
  \bibnamefont{and} \bibinfo{author}{\bibfnamefont{Z.}~\bibnamefont{Zhi}},
  \bibinfo{journal}{Chinese Physics Letter} \textbf{\bibinfo{volume}{27}},
  \bibinfo{pages}{087402} (\bibinfo{year}{2010}).

\bibitem[{\citenamefont{Durajski and
  Szcz{\c{e}}{\'s}niak}(2017)}]{Durajski2017A}
\bibinfo{author}{\bibfnamefont{A.~P.} \bibnamefont{Durajski}} \bibnamefont{and}
  \bibinfo{author}{\bibfnamefont{R.}~\bibnamefont{Szcz{\c{e}}{\'s}niak}},
  \bibinfo{journal}{Scientific Reports} \textbf{\bibinfo{volume}{7}},
  \bibinfo{pages}{4473} (\bibinfo{year}{2017}).

\bibitem[{\citenamefont{Szcz{\c{e}}{\'s}niak and
  Durajski}(2018)}]{Szczesniak2018A}
\bibinfo{author}{\bibfnamefont{R.}~\bibnamefont{Szcz{\c{e}}{\'s}niak}}
  \bibnamefont{and} \bibinfo{author}{\bibfnamefont{A.~P.}
  \bibnamefont{Durajski}}, \bibinfo{journal}{Scientific Reports}
  \textbf{\bibinfo{volume}{8}}, \bibinfo{pages}{6037} (\bibinfo{year}{2018}).

\bibitem[{\citenamefont{Peng et~al.}(2017)\citenamefont{Peng, Sun, Pickard,
  Needs, Wu, and Ma}}]{Peng2017A}
\bibinfo{author}{\bibfnamefont{F.}~\bibnamefont{Peng}},
  \bibinfo{author}{\bibfnamefont{Y.}~\bibnamefont{Sun}},
  \bibinfo{author}{\bibfnamefont{C.~J.} \bibnamefont{Pickard}},
  \bibinfo{author}{\bibfnamefont{R.~J.} \bibnamefont{Needs}},
  \bibinfo{author}{\bibfnamefont{Q.}~\bibnamefont{Wu}}, \bibnamefont{and}
  \bibinfo{author}{\bibfnamefont{Y.}~\bibnamefont{Ma}},
  \bibinfo{journal}{Physical Review Letters} \textbf{\bibinfo{volume}{119}},
  \bibinfo{pages}{107001} (\bibinfo{year}{2017}).

\bibitem[{\citenamefont{Miller et~al.}(1998)\citenamefont{Miller, Freericks,
  and Nicol}}]{Freericks1998A}
\bibinfo{author}{\bibfnamefont{P.}~\bibnamefont{Miller}},
  \bibinfo{author}{\bibfnamefont{J.~K.} \bibnamefont{Freericks}},
  \bibnamefont{and} \bibinfo{author}{\bibfnamefont{E.~J.} \bibnamefont{Nicol}},
  \bibinfo{journal}{Physical Review B} \textbf{\bibinfo{volume}{58}},
  \bibinfo{pages}{14498} (\bibinfo{year}{1998}).

\bibitem[{\citenamefont{Szcz{\c{e}}{\'s}niak}(2006)}]{Szczesniak2006A}
\bibinfo{author}{\bibfnamefont{R.}~\bibnamefont{Szcz{\c{e}}{\'s}niak}},
  \bibinfo{journal}{Acta Physica Polonica A} \textbf{\bibinfo{volume}{109}},
  \bibinfo{pages}{179} (\bibinfo{year}{2006}).

\bibitem[{\citenamefont{Kresin et~al.}(1993)\citenamefont{Kresin, Morawitz, and
  Wolf}}]{Kresin1993A}
\bibinfo{author}{\bibfnamefont{V.~Z.} \bibnamefont{Kresin}},
  \bibinfo{author}{\bibfnamefont{H.}~\bibnamefont{Morawitz}}, \bibnamefont{and}
  \bibinfo{author}{\bibfnamefont{S.~A.} \bibnamefont{Wolf}},
  \emph{\bibinfo{title}{Mechanisms of Conventional and High $T_{C}$
  Superconductivity}} (\bibinfo{publisher}{Oxford University Press},
  \bibinfo{year}{1993}).

\bibitem[{\citenamefont{Bardeen
  et~al.}(1957{\natexlab{a}})\citenamefont{Bardeen, Cooper, and
  Schrieffer}}]{Bardeen1957A}
\bibinfo{author}{\bibfnamefont{J.}~\bibnamefont{Bardeen}},
  \bibinfo{author}{\bibfnamefont{L.~N.} \bibnamefont{Cooper}},
  \bibnamefont{and} \bibinfo{author}{\bibfnamefont{J.~R.}
  \bibnamefont{Schrieffer}}, \bibinfo{journal}{Physical Review}
  \textbf{\bibinfo{volume}{106}}, \bibinfo{pages}{162}
  (\bibinfo{year}{1957}{\natexlab{a}}).

\bibitem[{\citenamefont{Bardeen
  et~al.}(1957{\natexlab{b}})\citenamefont{Bardeen, Cooper, and
  Schrieffer}}]{Bardeen1957B}
\bibinfo{author}{\bibfnamefont{J.}~\bibnamefont{Bardeen}},
  \bibinfo{author}{\bibfnamefont{L.~N.} \bibnamefont{Cooper}},
  \bibnamefont{and} \bibinfo{author}{\bibfnamefont{J.~R.}
  \bibnamefont{Schrieffer}}, \bibinfo{journal}{Physical Review}
  \textbf{\bibinfo{volume}{108}}, \bibinfo{pages}{1175}
  (\bibinfo{year}{1957}{\natexlab{b}}).

\bibitem[{\citenamefont{Szcz{\c{e}}{\'s}niak and
  Durajski}(2013{\natexlab{a}})}]{Szczesniak2013A}
\bibinfo{author}{\bibfnamefont{R.}~\bibnamefont{Szcz{\c{e}}{\'s}niak}}
  \bibnamefont{and} \bibinfo{author}{\bibfnamefont{A.~P.}
  \bibnamefont{Durajski}}, \bibinfo{journal}{Solid State Communications}
  \textbf{\bibinfo{volume}{172}}, \bibinfo{pages}{5}
  (\bibinfo{year}{2013}{\natexlab{a}}).

\bibitem[{\citenamefont{Duda et~al.}(2016)\citenamefont{Duda,
  Szcz{\c{e}}{\'s}niak, Sowi{\'n}ska, and Domagalska}}]{Duda2016A}
\bibinfo{author}{\bibfnamefont{A.~M.} \bibnamefont{Duda}},
  \bibinfo{author}{\bibfnamefont{R.}~\bibnamefont{Szcz{\c{e}}{\'s}niak}},
  \bibinfo{author}{\bibfnamefont{M.~A.} \bibnamefont{Sowi{\'n}ska}},
  \bibnamefont{and} \bibinfo{author}{\bibfnamefont{I.~A.}
  \bibnamefont{Domagalska}}, \bibinfo{journal}{Acta Physica Polonica A}
  \textbf{\bibinfo{volume}{130}}, \bibinfo{pages}{649} (\bibinfo{year}{2016}).

\bibitem[{\citenamefont{Szcz{\c{e}}{\'s}niak
  et~al.}(2010)\citenamefont{Szcz{\c{e}}{\'s}niak, Jarosik, and
  Szcz{\c{e}}{\'s}niak}}]{Szczesniak2010A}
\bibinfo{author}{\bibfnamefont{R.}~\bibnamefont{Szcz{\c{e}}{\'s}niak}},
  \bibinfo{author}{\bibfnamefont{M.~W.} \bibnamefont{Jarosik}},
  \bibnamefont{and}
  \bibinfo{author}{\bibfnamefont{D.}~\bibnamefont{Szcz{\c{e}}{\'s}niak}},
  \bibinfo{journal}{Physica B} \textbf{\bibinfo{volume}{405}},
  \bibinfo{pages}{4897} (\bibinfo{year}{2010}).

\bibitem[{\citenamefont{Durajski et~al.}(2016)\citenamefont{Durajski,
  Szcz{\c{e}}{\'s}niak, and Pietronero}}]{Durajski2016B}
\bibinfo{author}{\bibfnamefont{A.~P.} \bibnamefont{Durajski}},
  \bibinfo{author}{\bibfnamefont{R.}~\bibnamefont{Szcz{\c{e}}{\'s}niak}},
  \bibnamefont{and}
  \bibinfo{author}{\bibfnamefont{L.}~\bibnamefont{Pietronero}},
  \bibinfo{journal}{Annalen der Physik} \textbf{\bibinfo{volume}{528}},
  \bibinfo{pages}{358} (\bibinfo{year}{2016}).

\bibitem[{\citenamefont{Eremets et~al.}(2008)\citenamefont{Eremets, Trojan,
  Medvedev, Tse, and Yao}}]{Eremets2008A}
\bibinfo{author}{\bibfnamefont{M.~I.} \bibnamefont{Eremets}},
  \bibinfo{author}{\bibfnamefont{I.~A.} \bibnamefont{Trojan}},
  \bibinfo{author}{\bibfnamefont{S.~A.} \bibnamefont{Medvedev}},
  \bibinfo{author}{\bibfnamefont{J.~S.} \bibnamefont{Tse}}, \bibnamefont{and}
  \bibinfo{author}{\bibfnamefont{Y.}~\bibnamefont{Yao}},
  \bibinfo{journal}{Science} \textbf{\bibinfo{volume}{319}},
  \bibinfo{pages}{1506} (\bibinfo{year}{2008}).

\bibitem[{\citenamefont{Degtyareva et~al.}(2009)\citenamefont{Degtyareva,
  Proctor, Guillaume, Gregoryanz, and Hanfland}}]{Degtyareva2009A}
\bibinfo{author}{\bibfnamefont{O.}~\bibnamefont{Degtyareva}},
  \bibinfo{author}{\bibfnamefont{J.~E.} \bibnamefont{Proctor}},
  \bibinfo{author}{\bibfnamefont{C.~L.} \bibnamefont{Guillaume}},
  \bibinfo{author}{\bibfnamefont{E.}~\bibnamefont{Gregoryanz}},
  \bibnamefont{and} \bibinfo{author}{\bibfnamefont{M.}~\bibnamefont{Hanfland}},
  \bibinfo{journal}{Solid State Communications,}
  \textbf{\bibinfo{volume}{149}}, \bibinfo{pages}{1583} (\bibinfo{year}{2009}).

\bibitem[{\citenamefont{Allen and Dynes}(1975)}]{Allen1975A}
\bibinfo{author}{\bibfnamefont{P.~B.} \bibnamefont{Allen}} \bibnamefont{and}
  \bibinfo{author}{\bibfnamefont{R.~C.} \bibnamefont{Dynes}},
  \bibinfo{journal}{Physical Review B} \textbf{\bibinfo{volume}{12}},
  \bibinfo{pages}{905} (\bibinfo{year}{1975}).

\bibitem[{\citenamefont{Szcz{\c{e}}{\'s}niak and
  Durajski}(2013{\natexlab{b}})}]{Szczesniak2013B}
\bibinfo{author}{\bibfnamefont{R.}~\bibnamefont{Szcz{\c{e}}{\'s}niak}}
  \bibnamefont{and} \bibinfo{author}{\bibfnamefont{A.~P.}
  \bibnamefont{Durajski}}, \bibinfo{journal}{Solid State Sciences}
  \textbf{\bibinfo{volume}{25}}, \bibinfo{pages}{45}
  (\bibinfo{year}{2013}{\natexlab{b}}).

\bibitem[{\citenamefont{Fr{\"o}hlich}(1952)}]{Frohlich1952A}
\bibinfo{author}{\bibfnamefont{H.}~\bibnamefont{Fr{\"o}hlich}},
  \bibinfo{journal}{Proceedings of the Royal Society of London A}
  \textbf{\bibinfo{volume}{215}}, \bibinfo{pages}{291} (\bibinfo{year}{1952}).

\bibitem[{\citenamefont{Szcz{\c{e}}{\'s}niak}(2012)}]{Szczesniak2012A}
\bibinfo{author}{\bibfnamefont{R.}~\bibnamefont{Szcz{\c{e}}{\'s}niak}},
  \bibinfo{journal}{PloS ONE} \textbf{\bibinfo{volume}{7}},
  \bibinfo{pages}{e31873} (\bibinfo{year}{2012}).

\bibitem[{\citenamefont{Szcz{\c{e}}{\'s}niak
  et~al.}(2017)\citenamefont{Szcz{\c{e}}{\'s}niak, Durajski, and
  Duda}}]{Szczesniak2017A}
\bibinfo{author}{\bibfnamefont{R.}~\bibnamefont{Szcz{\c{e}}{\'s}niak}},
  \bibinfo{author}{\bibfnamefont{A.~P.} \bibnamefont{Durajski}},
  \bibnamefont{and} \bibinfo{author}{\bibfnamefont{A.~M.} \bibnamefont{Duda}},
  \bibinfo{journal}{Annalen der Physik} \textbf{\bibinfo{volume}{529}},
  \bibinfo{pages}{1600254} (\bibinfo{year}{2017}).

\bibitem[{\citenamefont{Ashcroft}(2004)}]{Ashcroft2004A}
\bibinfo{author}{\bibfnamefont{N.~W.} \bibnamefont{Ashcroft}},
  \bibinfo{journal}{Physical Review Letters} \textbf{\bibinfo{volume}{92}},
  \bibinfo{pages}{187002} (\bibinfo{year}{2004}).

\bibitem[{\citenamefont{Bennemann and Garland}(1972)}]{Bennemann1972A}
\bibinfo{author}{\bibfnamefont{K.}~\bibnamefont{Bennemann}} \bibnamefont{and}
  \bibinfo{author}{\bibfnamefont{J.}~\bibnamefont{Garland}},
  \bibinfo{journal}{American Institute of Physics Conference Proceedings}
  \textbf{\bibinfo{volume}{4}}, \bibinfo{pages}{103} (\bibinfo{year}{1972}).

\bibitem[{\citenamefont{Szcz{\c{e}}{\'s}niak and
  Drzazga}(2013)}]{Szczesniak2013C}
\bibinfo{author}{\bibfnamefont{R.}~\bibnamefont{Szcz{\c{e}}{\'s}niak}}
  \bibnamefont{and} \bibinfo{author}{\bibfnamefont{E.~A.}
  \bibnamefont{Drzazga}}, \bibinfo{journal}{Solid State Sciences}
  \textbf{\bibinfo{volume}{19}}, \bibinfo{pages}{167} (\bibinfo{year}{2013}).

\bibitem[{\citenamefont{McMillan}(1968)}]{McMillan1968A}
\bibinfo{author}{\bibfnamefont{W.~L.} \bibnamefont{McMillan}},
  \bibinfo{journal}{Physical Review} \textbf{\bibinfo{volume}{167}},
  \bibinfo{pages}{331} (\bibinfo{year}{1968}).

\bibitem[{\citenamefont{Wiendlocha et~al.}(2016)\citenamefont{Wiendlocha,
  Szcz{\c{e}}{\'s}niak, Durajski, and Muras}}]{Wiendlocha2016A}
\bibinfo{author}{\bibfnamefont{B.}~\bibnamefont{Wiendlocha}},
  \bibinfo{author}{\bibfnamefont{R.}~\bibnamefont{Szcz{\c{e}}{\'s}niak}},
  \bibinfo{author}{\bibfnamefont{A.~P.} \bibnamefont{Durajski}},
  \bibnamefont{and} \bibinfo{author}{\bibfnamefont{M.}~\bibnamefont{Muras}},
  \bibinfo{journal}{Physical Review B} \textbf{\bibinfo{volume}{94(13)}},
  \bibinfo{pages}{134517} (\bibinfo{year}{2016}).

\bibitem[{\citenamefont{Eschrig}(2001)}]{Eschrig2001A}
\bibinfo{author}{\bibfnamefont{H.}~\bibnamefont{Eschrig}},
  \emph{\bibinfo{title}{Theory of Superconductivity: A Primer}}
  (\bibinfo{publisher}{Citeseer}, \bibinfo{year}{2001}).

\bibitem[{\citenamefont{Beach et~al.}(2000)\citenamefont{Beach, Gooding, and
  Marsiglio}}]{Beach2000A}
\bibinfo{author}{\bibfnamefont{K.~S.~D.} \bibnamefont{Beach}},
  \bibinfo{author}{\bibfnamefont{R.~J.} \bibnamefont{Gooding}},
  \bibnamefont{and}
  \bibinfo{author}{\bibfnamefont{F.}~\bibnamefont{Marsiglio}},
  \bibinfo{journal}{Physical Review B} \textbf{\bibinfo{volume}{61}},
  \bibinfo{pages}{5147} (\bibinfo{year}{2000}).

\bibitem[{\citenamefont{Bardeen and Stephen}(1964)}]{Bardeen1964A}
\bibinfo{author}{\bibfnamefont{J.}~\bibnamefont{Bardeen}} \bibnamefont{and}
  \bibinfo{author}{\bibfnamefont{M.}~\bibnamefont{Stephen}},
  \bibinfo{journal}{Physical Review} \textbf{\bibinfo{volume}{136}},
  \bibinfo{pages}{A1485} (\bibinfo{year}{1964}).

\bibitem[{\citenamefont{Rose-Innes and Rhoderick}(1992)}]{Rose-Innes1969A}
\bibinfo{author}{\bibfnamefont{A.~C.} \bibnamefont{Rose-Innes}}
  \bibnamefont{and} \bibinfo{author}{\bibfnamefont{E.~H.}
  \bibnamefont{Rhoderick}}, \emph{\bibinfo{title}{Introduction to
  superconductivity}} (\bibinfo{publisher}{Pergamon Press},
  \bibinfo{year}{1992}).

\end{thebibliography}
\end{document}